\documentclass[10pt,journal,compsoc]{IEEEtran}
\ifCLASSOPTIONcompsoc
  \usepackage[nocompress]{cite}
\else
  \usepackage{cite}
\fi
\ifCLASSINFOpdf
\else
\fi

\usepackage{cite}
\usepackage{amsmath,amssymb,amsfonts}
\usepackage{algorithm}
\usepackage{algorithmic}
\usepackage{graphicx}
\usepackage{textcomp}
\usepackage{makecell}
\usepackage[dvipsnames]{xcolor}
\def\BibTeX{{\rm B\kern-.05em{\sc i\kern-.025em b}\kern-.08em
    T\kern-.1667em\lower.7ex\hbox{E}\kern-.125emX}}

\usepackage{adjustbox}
\usepackage{graphicx,amssymb,lineno}
\usepackage{multirow}
\usepackage[tight,footnotesize]{subfigure}
\usepackage{epsfig}
\usepackage{footmisc}
\usepackage{footnpag}
\usepackage{footnote}
\usepackage{hyperref}

\usepackage[inline]{enumitem}
\usepackage[group-digits=integer, group-separator={,}]{siunitx}

\usepackage{mathtools}

\DeclarePairedDelimiter\abs{\lvert}{\rvert}
\DeclarePairedDelimiter\norm{\lVert}{\rVert}

\usepackage{xparse}

\NewDocumentCommand\N{sm}{\mathcal{N}\IfBooleanT{#1}{^{\ast}}_{#2}}

\makeatletter
\newcommand\@ttstyle[1]{\text{\ttfamily\upshape #1}}

\newcommand\InputArray{\@ttstyle{u}}
\newcommand\at[1]{\@ttstyle{\lbrack}#1\@ttstyle{\rbrack}}

\newcommand\uMC{\@ttstyle{u\char`_mc}}
\newcommand\newMC{\@ttstyle{\~{u}\char`_mc}}
\newcommand\cost{\@ttstyle{cost}}
\newcommand\round{\@ttstyle{round}}
\makeatother

\newcommand\rr[1]{\tilde{#1}}

\newtheorem{theorem}{Theorem}
\newtheorem{lemma}[theorem]{Lemma}

\begin{document}
%
\title{MGARD+: Optimizing Multilevel Methods for Error-bounded Scientific Data Reduction}
%
%
%
%

\author{
Xin Liang, 
Ben Whitney, 
Jieyang Chen, 
Lipeng Wan, 
Qing Liu,
Dingwen Tao,
James Kress, \\
David Pugmire, 
Matthew Wolf, 
Norbert Podhorszki, and 
Scott Klasky, Senior Member, IEEE
\IEEEcompsocitemizethanks{\IEEEcompsocthanksitem X. Liang, B. Whitney, J. Chen, L. Wan, J. Kress, D. Pugmire, M. Wolf, N. Podhorszki, S. Klasky are with the Computer Science and Mathematics Division, Oak Ridge National Laboratory, Oak Ridge, TN 37830.\protect\\
E-mail: \{liangx, whitneybe, chenj3, wanl, kressjm, pugmire, wolfmd, pnorbert, klasky\}@ornl.gov
\IEEEcompsocthanksitem Q. Liu is with the Helen and John C. Hartmann Department of Electrical and Computer Engineering, New Jersey Institute of Technology, Newark, NJ 07102.\protect\\
E-mail: qliu@njit.edu
\IEEEcompsocthanksitem D. Tao is with the School of Electrical Engineering \& Computer Science, Washington State University, Pullman, WA 99163.\protect\\
E-mail: dingwen.tao@wsu.edu
}
\thanks{
\textit{X. Liang and B. Whitney were co-first authors of this work. This work has been submitted to the IEEE for possible publication. Copyright may be transferred without notice, after which this version may no longer be accessible.}}
}

%
%

\markboth{Journal of \LaTeX\ Class Files,~Vol.~14, No.~8, August~2015}%
{Shell \MakeLowercase{\textit{et al.}}: Bare Demo of IEEEtran.cls for Computer Society Journals}
%



\IEEEtitleabstractindextext{%
\begin{abstract}
Data management is becoming increasingly important in dealing with the large amounts of data produced by today's large-scale scientific simulations and instruments. 
Existing multilevel compression algorithms offer a promising way to manage scientific data at scale, but may suffer from relatively low performance and reduction quality. 
In this paper, we propose MGARD+, a multilevel data reduction and refactoring framework drawing on previous multilevel methods, to achieve high-performance data decomposition and high-quality error-bounded lossy compression.
Our contributions are four-fold:
\begin{enumerate*}
\item We propose a level-wise coefficient quantization method, which uses different error tolerances to quantize the multilevel coefficients.
\item We propose an adaptive decomposition method which treats the multilevel decomposition as a preconditioner and terminates the decomposition process at an appropriate level. 
\item We leverage a set of algorithmic optimization strategies to significantly improve the performance of multilevel decomposition/recomposition. 
\item We evaluate our proposed method using four real-world scientific datasets and compare with several state-of-the-art lossy compressors.
\end{enumerate*}
Experiments demonstrate that our optimizations improve the decomposition/recomposition performance of the existing multilevel method by up to \(70 \times\), and the proposed compression method can improve compression ratio by up to \(2 \times\) compared with other state-of-the-art error-bounded lossy compressors under the same level of data distortion.
\end{abstract}

}

\maketitle

\IEEEdisplaynontitleabstractindextext

%
\IEEEpeerreviewmaketitle

\IEEEraisesectionheading{\section{Introduction}\label{sec:introduction}}
\IEEEPARstart{W}{ith} the extreme amounts of data produced by large-scale scientific simulations on leadership high-performance computing (HPC) systems and scientific instruments, data management has become a serious problem. 
On the one hand, not all the data can be stored in the parallel file systems. 
They have to be moved to relatively slow storage devices such as archives, where the data transfer time will become prohibitive because of the limited I/O bandwidth.
On the other hand, even if the full data could be stored, post hoc analysis on the entire data would be too costly to conduct.
For instance, the DCA++ code~\cite{summers2009dca++} produces 100~TB of data in a single run, but only a small subset (100~MB) is written in order to lower the cost of post hoc analysis. In particular, scientists reduce their seven dimensional tensor down to a two dimensional subset, thus reducing the data by a factor of $10^6$.
This operation makes it possible for scientists to conduct data analysis on a laptop, but valuable information not captured in the reduced data 
set is often lost.

Error-bounded lossy compression techniques~\cite{sz16, zfp, fpzip, isabela, numarck, sz-reg, sz-pwr, sz-hybrid} have been proposed and developed in the last decade to address the storage issue.
These techniques aim to significantly reduce data size while controlling distortion in the decompressed data. 
However, the methods suffer from large distortion when the required compression ratio is relatively high (e.g., $30\times$) -- a common demand for data-intensive HPC applications. 
For example, ZFP~\cite{zfp} exhibits visual artifacts when the compression ratio reaches $64\times$ according to previous studies \cite{sz-reg}.
SZ~\cite{sz-reg} generates visually better results thanks to its multi-algorithm design, but it may still cause undesired data distortion as the compression ratio is increased to relatively high levels.
The hybrid model proposed in~\cite{sz-hybrid} significantly improves the compression quality by integrating ZFP's orthogonal transform in the SZ compression framework.
This integration comes at a significant computational cost, though, and in practice a qualified compressor must achieve not only high compression ratios but also high compression and decompression speeds. 

Recently, the applied math community has proposed a new method, MultiGrid Adaptive Reduction of Data (MGARD)~\cite{ainsworth2018multilevel, ainsworth2019multilevel, ainsworth2019multilevelerror}, for compression of scientific data.
Drawing on the theories of wavelet analysis, finite element methods, and multigrid linear solvers, MGARD decomposes multidimensional datasets into a collection of components of varying scale and resolution.
The coefficients for these components are then adaptively quantized to achieve error-bounded compression.
Like multigrid linear solvers, MGARD makes use of a sequence of nested grids to achieve a separation of scales, and adapts its treatment of each component to that component's scale.
It is not, however, a linear solver, and the wide variety of improvements made to the multigrid method are not immediately transferable to the domain of data compression.
As an error-bounded lossy compressor, MGARD is unique for providing strict error control for derived quantities~\cite{ainsworth2019multilevelerror}, 
in addition to the point-wise error control provided by existing approaches.

Besides data reduction, multilevel decomposition as implemented in MGARD can also be used for data refactoring.
Both data reduction and data refactoring aim to shrink large input datasets, but they differ in how the compressed representations can subsequently be used.
The output of a general lossy data reduction method can only be decompressed in full resolution, resulting in a lossy reconstruction having the same size as the original input.
The output of a data refactoring method, by contrast, can be partially decompressed to produce reconstructions of intermediate accuracy and size.
This may be accomplished, by example, by decomposing the original data into a hierarchical sequence of components, a subset of which may be summed to produce a reconstruction comprising fewer degrees of freedom than the original input.
Post hoc analysis may then be carried out on this reduced representation, at a reduced cost.
This is especially valuable if, for example, the original input dataset is too large to be analyzed by the available computational resources (a laptop, rather than a cluster, for instance).
Although data refactoring methods may incur larger distortion at a given compression ratio than data reduction methods, they have two particular advantages:
\begin{enumerate*}
\item they allow for progressive reconstruction of data, with precision improving as more storage space is allocated, and
\item they can be used to generate coarse-grained representations on which post hoc data analysis may be performed with greatly reduced computational complexity.
\end{enumerate*}

The main contributions of this paper is to design and evaluate tailored optimizations to MGARD, which, despite offering an elegant approach to the problem of scientific data reduction, suffers from fairly low throughput and suboptimal compression ratios.
Specifically, we propose two novel approaches to improve compression ratio at a given level of distortion. 
In addition, we develop a series of optimization strategies for the multilevel methods, which can substantially boost the performance of both data decomposition and recomposition. 
Such optimizations are also important for data refactoring use cases, as will be demonstrated in our evaluation. 
In summary, our contributions are as follows:
\begin{itemize}
    \item \textit{Adaptive error-based coefficient quantization}: We propose to use different error tolerances to quantize the multilevel coefficients at each level in the previous multilevel algorithm, which can significantly improve the compression ratios under given distortions in terms of Peak Signal-to-Noise Ratio (PSNR). To this end, we carefully analyze the impact of the quantization method and choose the best-fit scaling factor determining the relationship of error tolerance across different levels.
    \item \emph{Adaptive data decomposition termination}: We propose to treat the multilevel data decomposition as a preconditioner, unlike the traditional multilevel compressor, MGARD, which totally relies on the data decomposition for the whole compression procedure. In our approach, the multilevel decomposition would terminate at an appropriate level, and the remaining coarse-grained representation is compressed via external error-bounded lossy compressors. This can further improve compression ratios over the first strategy. 
    \item \textit{A series of performance optimizations}: With algorithmic improvements, we significantly improve the performance of our error-bounded lossy compression method over the traditional baseline. Specifically, we adopt a level-centric data reordering strategy and batched operations to improve cache coherence and memory efficiency. We also revise the correction computation kernel (one of the most important steps in the multilevel decomposition/recomposition algorithms) to reduce computational cost.
    \item \textit{Thorough evaluation}: We evaluate our method with respect to both performance and quality using four real-world datasets from different scientific applications. We first demonstrate the effectiveness of the proposed optimizations compared to original multilevel approach~\cite{ainsworth2019multilevel}, and then compare our method to state-of-the-art error-bounded lossy compressors including SZ~\cite{sz-reg}, ZFP~\cite{zfp}, and the hybrid model~\cite{sz-hybrid}. 
    Experiments show that our proposed method has a $20\sim70\times$ performance improvement over the previous multilevel method in terms of decomposition/recomposition speed, and the evaluation results on the iso-surface mini-analysis indicate that conducting scientific analysis on the coarse-grained representations could significantly improve the analysis performance. Furthermore, our solution yields $2 \times$ compression ratio improvement over state-of-the-art error-bounded lossy compressors~\cite{sz-reg, zfp, sz-hybrid} at the same distortion, especially in the high compression ratio cases, showing great potential in reducing the storage overhead. 
\end{itemize}

The rest of the paper is organized as follows. 
In Section~\ref{sec:background}, we summarize the main operations of MGARD.
In Section~\ref{sec:formulation}, we specify the metrics for evaluation. 
In Section~\ref{sec:reduction}, we propose level-wise quantization and adaptive decomposition to improve the quality of error-bounded lossy compression. 
In Section~\ref{sec:optimization}, we introduce a set of optimization techniques applied to our implementation.
In Section~\ref{sec:evaluation}, we evaluate our method using real-world simulation data from scientific applications. 
Finally, we discuss related work in Section~\ref{sec:related} and conclude with a vision for future work in Section~\ref{sec:conclusion}.
\section{Background}\label{sec:background}

In this section, we describe the central decomposition and recomposition routines of MGARD, which serves as the starting point for the multilevel method in this work.
This description focuses on the computational steps involved.
For a full mathematical treatment, see \cite{ainsworth2018multilevel, ainsworth2019multilevel}. 
Some frequently used symbols are summarized in Table~\ref{tab:notation}.

\begin{table}[t]
\scriptsize
\centering
\caption{Description of frequently used symbols}
\vspace{-2mm}
\label{tab:notation}
\begin{tabular}{|c|l|}
\hline
\thead{Symbol} &  \thead{Description}\\ \hline
\(u\) & Input data array.\\ \hline
$\Tilde{u}$ & Reconstructed data array. \\ \hline
$\uMC$ & Multilevel coefficients.\\ \hline
$\newMC$ & Quantized multilevel coefficients.\\ \hline
$L$ & Maximum multilevel decomposition level.\\ \hline
$d$ & Spatial dimension.\\ \hline
$n_i$ & Number of elements along the $i$-th dimension.\\ \hline
$\N{l}$ & Level $l$ subgrids (nodal nodes in level $l + 1$).\\ \hline
$\N*{l}$ & Level $l$ displaced nodes (coefficient nodes).\\ \hline
$Q_l$ & The $L^2$ projection operator.\\ \hline
$\Pi_l$ & The piecewise multilinear interpolation operator.\\ \hline
$I$ & Identity operator.\\ \hline
$h_l$ & Internode spacing in level $l$.\\ \hline
$\tau$ & User-specified error tolerance.\\ \hline
$\kappa$ & Scaling factor for level-wise quantization.\\ \hline
$C_{L^{2}}$ & Derived constant in~\cite{ainsworth2019multilevelerror} for $L^2$ error guarantee.\\ \hline
$C_{L^{\infty}}$ & Derived constant in~\cite{ainsworth2019multilevel} for $L^\infty$ error guarantee.\\ \hline
\end{tabular}
\vspace{-4mm}
\end{table}

The input is an array (multidimensional in general) \(\InputArray\) of floating-point numbers.
We interpret \(\InputArray\) as the values taken by a function \(u\) on a grid \(\N{L}\) having the same dimensions as \(\InputArray\).
For example, if \(\InputArray\) has shape \(n_{1} \times \dotsb \times n_{d}\), then \(\N{L}\) might be \(\{(j_{1} h, \dotsc, j_{d} h) : 0 \leq j_{i} < n_{i}\}\).
Each element \(x \in \N{L}\) is a point in the domain of \(u\); the corresponding entry \(\InputArray\at{(j_{1}, \dotsc, j_{d})}\) of the array is the value \(u(x)\) taken by the function at that point.

\begin{figure}[ht] \centering
\vspace{-3mm}
\includegraphics[width=\columnwidth]{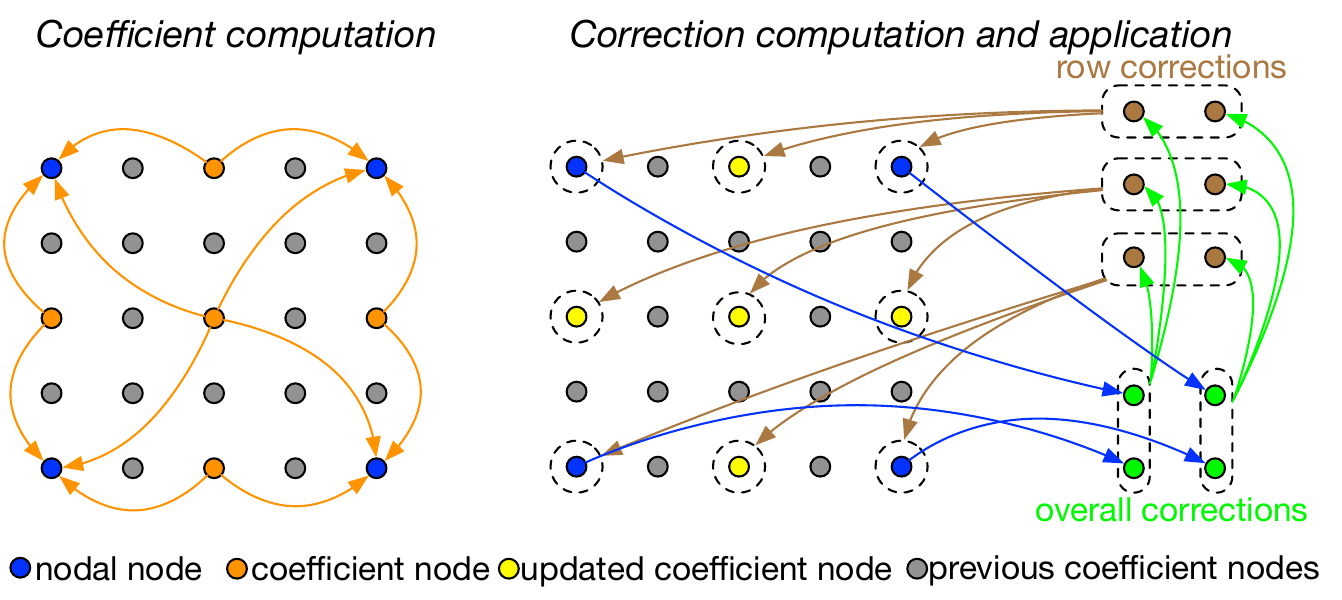}
\vspace{-6mm}
\caption{Data layout in MGARD and the strided dependencies ($l = L-1$).}
\label{fig:dependency}
\vspace{-2mm}
\end{figure}

We decompose \(\InputArray\) using a sequence \(\N{L - 1}, \dotsc, \N{0}\) of subgrids of \(\N{L}\).
We require that the sequence be decreasing, i.e., that \(\N{l + 1} \supset \N{l}\). 
See Fig.~\ref{fig:dependency} for an example.
The blue nodes comprise \(\N{l}\), the blue and orange nodes comprise \(\N{l + 1}\), and the blue, orange, and grey nodes comprise \(\N{l + 2}\).
Denote by \(\N*{l}\) the set \(\N{l} \setminus \N{l - 1}\), with \(\N{-1} = \emptyset\).
We define for \(0 \leq l \leq L\) operators \(Q_{l}\) and \(\Pi_{l}\), each outputting an array of values defined on \(\N{l}\).
\(Q_{l}\) is an \(L^{2}\) projection operator.
It is applied by computing a matrix--vector product and then solving a linear system.
\(\Pi_{l}\) is a multilinear interpolation operator.
It leaves values on \(\N{l - 1}\) unchanged; values on \(\N*{l}\) are determined by interpolating values on \(\N{l - 1}\).
Mathematically, we can interpret the arrays output by these operators as functions in appropriately defined function spaces.
See \cite{ainsworth2018multilevel, ainsworth2019multilevel} for details. 
For a given level $l$, we refer to nodes in \(\N{l - 1}\) as \emph{nodal nodes} and to those in \(\N*{l}\) as \emph{coefficient nodes}.

The decomposition routine transforms the input function \(u\) to the \emph{multilevel components} \(\{(I - \Pi_{l - 1})Q_{l} u : 0 \leq l \leq L\}\).
This is accomplished by the following iterative procedure, with \(l = L\) for the first iteration.
\begin{enumerate}
\item Start with \(Q_{l} u\), with \(Q_{L} u = u\).
\item\label{alg:decomposition:itm:InterpolantComputation} Compute the interpolant \(\Pi_{l - 1} Q_{l} u\).
\item\label{alg:decomposition:itm:InterpolationSubtraction} Subtract the interpolant \(\Pi_{l - 1} Q_{l} u\) from \(Q_{l} u\), obtaining the multilevel component \((I - \Pi_{l - 1})Q_{l} u\).
\item\label{alg:decomposition:itm:CorrectionComputation} Compute the projection \(Q_{l - 1}(I - \Pi_{l - 1})Q_{l} u\) of the multilevel component.
It can be shown that this projection, which we call the \emph{correction}, is equal to \(Q_{l - 1} u - \Pi_{l - 1} Q_{l} u\).
\item\label{alg:decomposition:itm:CorrectionApplication} Add the correction \(Q_{l - 1} u - \Pi_{l - 1} Q_{l} u\) to the interpolant \(\Pi_{l - 1} Q_{l} u\), obtaining \(Q_{l - 1} u\).
\item If \(l = 0\), stop.
Otherwise, decrement \(l\) and repeat.
\end{enumerate}
We call Step~\ref{alg:decomposition:itm:CorrectionComputation} the \emph{correction computation} and Step~\ref{alg:decomposition:itm:CorrectionApplication} the \emph{correction application}.
Steps~\ref{alg:decomposition:itm:InterpolantComputation} and~\ref{alg:decomposition:itm:InterpolationSubtraction} are combined in the implementation into a single operation which we call \emph{coefficient computation}.
The coefficients in question are the nodal values \(\{(I - \Pi_{l - 1})Q_{l} u(x) : x \in \N*{l}\}\).
The output of the decomposition routine is an array containing these coefficients for each level \(l\).
We call these values the \emph{multilevel coefficients} and denote the collection \(\uMC\).
\(\uMC\) is indexed by the nodes of the finest grid \(\N{L}\): given \(x \in \N{L}\), if \(l\) is the least grid index such that \(x \in \N{l}\), then \(\uMC\at{x} = (I - \Pi_{l - 1})Q_{l} u(x)\).
Following decomposition, the next step in MGARD is to quantize \(\uMC\).
See Section~\ref{subsec:quantization}.

The recomposition procedure is the inverse of the decomposition procedure.
We start with multilevel components \(\{(I - \Pi_{l - 1})Q_{l} u : 0 \leq l \leq L\}\).
We recover \(u\) by the following iterative procedure, starting with \(l = 0\).
\begin{enumerate}
\item Start with \(Q_{l - 1} u\), with \(Q_{-1} u = 0\).
\item\label{alg:recomposition:itm:CorrectionComputation} Compute the projection \(Q_{l - 1} (I - \Pi_{l - 1}) Q_{l} u\) of the multilevel component.
As in the decomposition routine, we call this projection the \emph{correction}.
It is again equal to \(Q_{l - 1} u - \Pi_{l - 1} Q_{l} u\).
\item\label{alg:recomposition:itm:CorrectionApplication} Subtract the correction \(Q_{l - 1} u - \Pi_{l - 1} Q_{l} u\) from \(Q_{l - 1} u\), obtaining the interpolant \(\Pi_{l - 1} Q_{l} u\).
\item\label{alg:recomposition:itm:ProjectionCancellation} Add the interpolant \(\Pi_{l - 1} Q_{l} u\) to the multilevel component \((I - \Pi_{l - 1})Q_{l} u\), obtaining \(Q_{l} u\).
\item If \(l = L\), stop.
Otherwise, increment \(l\) and repeat.
\end{enumerate}
The recomposition and decomposition procedures require a very similar set of subroutines.
The correction is computed in Step~\ref{alg:recomposition:itm:CorrectionComputation} and applied in Step~\ref{alg:recomposition:itm:CorrectionApplication}, though it is subtracted rather than added here.
Step~\ref{alg:recomposition:itm:ProjectionCancellation} is a simple inverse of the coefficient computation.
The hierarchical nature of the multilevel algorithm leads to strided memory access in these operations, as seen in Fig.~\ref{fig:dependency}.
An approach to mitigate this problem is detailed in Section~\ref{subsec:reorder}.
\section{Metrics}\label{sec:formulation}
In this paper, we focus on improving the performance and quality of the previous multilevel method~\cite{ainsworth2019multilevel}.
We briefly introduce our metrics for the two objectives in this section.

\vspace{-3mm}
\subsection{Performance}
We measure the performance of multilevel operations in terms of throughput, which is evaluated by $\text{\ttfamily size} / t$, where $\text{\ttfamily size}$ is the original data size and $t$ is the time used for the operation (such as decomposition, recomposition, compression, or decompression).
The overall throughput on a dataset (which may contain multiple fields, each operated on separately) is computed by dividing the total size by the total time.

\vspace{-3mm}
\subsection{Quality}
We measure the quality of data reduction using rate--distortion graphs~\cite{berger2003rate}, which give a visual representation of how much compression can be achieved using lossy compression methods. 
The rate (a.k.a.\ the bit-rate), on the $X$ axis of the graph, is the average number of bits per data point in the reduced representation.
It is equal to the number of bits per field in the original dataset divided by the compression ratio.
As mentioned before, we adopt PSNR as the distortion metric, on the $Y$ axis of the graph, because it is equivalent to the commonly used statistical error and often serves as an indicator of visual quality. PSNR has been widely used in much previous work, including~\cite{sz16, sz17, sz-reg, zfp, sz-hybrid}. 
PSNR is computed as follows:
\vspace{-1mm}
\begin{align*}
\vspace{-1mm}
\text{PSNR} &= 20 \log_{10}(\max(u_i) - \min(\Tilde{u}_i))\\
&\qquad {}- 10 \log_{10}(\textstyle\sum_{i=1}^{N}{(u_i - \Tilde{u}_i)^2}/N)
\end{align*}
where $\{u_1, u_2, \hdots, u_N\}$ and $\{\Tilde{u}_1, \Tilde{u}_2, \hdots, \Tilde{u}_N\}$ are the original and decompressed data, respectively ($N$ is the number of data points).
Higher PSNR indicates less error, and thus higher quality. 
So, our target in improving data reduction quality turns out to be maximizing the PSNR of the decompressed data at a fixed compression ratio, or, conversely, maximizing the compression ratio at a fixed PSNR. 
As PSNR is computed over the sum of squared differences between the original and decompressed data (i.e., $\frac{1}{N} \sum_{i=1}^{N}{(u_i - \Tilde{u}_i)^2}$), we focus on minimizing the squared $L^2$ norm in our analysis. 
\section{Multilevel reduction with level-wise quantization and adaptive decomposition}\label{sec:reduction}
We propose two novel techniques to improve the compression ratios of the multilevel method at the same PSNR. 
First, in Section~\ref{subsec:quantization}, a level-wise quantization method significantly improves the compression ratios when error tolerance is high.
Second, in Section~\ref{subsec:decomposition}, an adaptive decomposition method automatically terminates the multilevel decomposition process at an appropriate level and compresses the remaining coarse-grained representation with external compressors, which further improves the compression ratios.

\vspace{-1mm}
\subsection{Level-wise quantization}\label{subsec:quantization}
As detailed in Section~\ref{sec:background}, the first step in MGARD's compression stage is the decomposition of the input \(u\) into a set of multilevel coefficients \(\uMC\).
Each entry of \(\uMC\) is then quantized, yielding a quantized set of multilevel coefficients \(\newMC\).
Just as \(\uMC\) encodes the input function \(u\), \(\newMC\) encodes an approximation \(\rr{u}\) to \(u\).
Care must be taken when quantizing that the error \(\norm{u - \rr{u}}_{L^{2}}\) between the two is not greater than the error tolerance \(\tau_{L^{2}}\) prescribed by the user.
It is therefore useful to relate the individual quantization errors \(\uMC\at{x} - \newMC\at{x}\) to the overall approximation error \(u - \rr{u}\).
Such a relationship is given in~\cite{ainsworth2019multilevelerror}, where it is proved that, in the case of uniform grids, for \(C_{L^{2}}\) a constant depending on the grid hierarchy,
\vspace{-1mm}
\begin{equation*}
\vspace{-1mm}
\begin{split}
&\text{if} \quad \sum_{l = 0}^{L} \sum_{x \in \N*{l}} {h_{l}^{d} \abs{\uMC\at{x} - \newMC\at{x}}^{2}} \leq \frac{\tau_{L^{2}}^{2}}{C_{L^{2}}}
,\\
&\text{then} \quad \norm{u - \rr{u}}_{L^{2}} \leq \tau_{L^{2}}
.
\end{split}
\label{eqn:L2NormEstimator}
\end{equation*}
Here \(d\) is the spatial dimension and \(h_{l}\) is the spacing between nodes in \(\N{l}\) (assumed to be uniform across dimensions).
In view of this condition, the quantizer has an error `budget' of \(\tau_{L^{2}} / C_{L^{2}}^{1 / 2}\) to be distributed among the \(L + 1\) levels \(\{\uMC\at{x} : x \in \N*{l}\}\).

The next task is to quantize each coefficient \(\uMC\at{x}\).
This can be accomplished by splitting the range of the multilevel coefficients into labelled bins of uniform width \(q\).
\(\uMC\at{x}\) can then be mapped to the label of the bin containing it.
That label encodes \(\newMC\at{x}\).
If we choose \(\newMC\at{x}\) to be the center of the bin, then \(\abs{\uMC\at{x} - \newMC\at{x}} \leq q / 2\), since the bin has width \(q\) and contains \(\uMC\at{x}\).
If the range of the multilevel coefficients has size \(R = \max(\uMC) - \min(\uMC)\), then \(\lceil R / q\rceil\) labels suffice to quantize all of the multilevel coefficients with error at most \(q / 2\).
After quantization, the labels are passed to a lossless encoder for compression.

How many bits are required to store the labels produced by quantization?
The exact answer will depend on the distribution of the multilevel coefficients and the lossless encoder used, but we can estimate the cost using Shannon entropy~\cite{shannon2001mathematical}.
Informally and in brief, the cost in bits per symbol to encode a stream is bounded below by the Shannon entropy of the source, and this rate can be matched asymptotically.
Here, where the symbol alphabet has size \(\lceil R / q\rceil\), the entropy is at most \(\log_{2}(\lceil R / q\rceil)\).
So, for quantization with bin width \(q\), the cost in bits to store the quantized coefficients can be estimated by \(\# \N{L} \log_{2}(R / q)\) if the same bin width is used for all coefficients.

We seek to improve on this cost by more effectively distributing the error `budget.'
To do this, we quantize the coefficients separately by level.
We call this strategy \emph{level-wise quantization}.
Suppose the coefficients \(\{\uMC\at{x} : x \in \N*{l}\}\) are quantized with bin width \(q_{l}\).
Then
\vspace{-1mm}
\begin{equation*}
\vspace{-1mm}
\sum_{x \in \N*{l}} h_{l}^{d} \abs{\uMC\at{x} - \newMC\at{x}}^{2} \leq h_{l}^{d} \#\N*{l} (q_{l} / 2)^{2}
\label{eqn:SingleLevelSquareQuantizationError}
\end{equation*}
and the estimated cost of encoding the quantization labels for the level is \(\#\N*{l} \log_{2}(R / q_{l})\).
Summing over \(l\), we define an estimated cost function \(\cost\) by
\vspace{-2mm}
\begin{equation*}
\vspace{-1.5mm}
\cost(q_{0}, \dotsc, q_{L}) = \sum_{l = 0}^{L} \#\N*{l} \log_{2}(R / q_{l})
.
\end{equation*}
Consider the optimization problem
\vspace{-1mm}
\begin{align*}
\vspace{-2mm}
&\text{minimize} \quad \cost(q_{0}, \dotsc, q_{L})\\
\vspace{-2mm}
&\text{subject to} \quad \sum_{l = 0}^{L} h_{l}^{d} \#\N*{l} (q_{l} / 2)^{2} = \frac{\tau_{L^{2}}^{2}}{C_{L^{2}}}
.
\vspace{-1mm}
\end{align*}
(The cost decreases as the bin widths increase, so nothing is lost by making the error condition constraint an equality.)
A straightforward application of Lagrange multipliers and the convexity of \(\cost\), which we omit to save space, shows that the solution this problem is given by \(q_{l} = 2 \tau_{L^{2}} / (C_{L^{2}} h_{l}^{d}  \#\N{L})^{1 / 2}\).
The estimated cost of this quantization strategy is
\vspace{-2mm}
\begin{equation*}
\vspace{-2mm}
\cost(q_{0}, \dotsc, q_{L})
= \sum_{l = 0}^{L} \#\N*{l} \log_{2}\Biggl(\frac{R \sqrt{C_{L^{2}} h_{l}^{d} \#\N{L}}}{2 \tau_{L^{2}}}\Biggr)
.
\end{equation*}

For ease of notation in subsequent sections, it will be convenient to describe our quantization strategy in terms of quantization error tolerances, rather than quantization bin widths, for each level.
That is, we will quantize \(\{\uMC\at{x} : x \in \N*{l}\}\) so that
\vspace{-1mm}
\begin{equation*}
\vspace{-1mm}
\max_{x \in \N*{l}} {\abs{\uMC\at{x} - \newMC\at{x}}} \leq \tau_{l}
\end{equation*}
for some quantization error tolerance \(\tau_{l}\).
The bin widths \(q_{l}\) found above correspond to \(\tau_{l} = \tau_{L^{2}} / (C_{L^{2}} h_{l}^{d} \#\N{L})^{1 / 2}\).
With an archetypical grid hierarchy used by MGARD, the grid resolution doubles in each dimension from one level to the next, and so \(h_{l} \simeq 2^{-l}\).
Then the quantization error tolerance grows from one level to the next by a factor of
\vspace{-2mm}
\begin{equation*}
\vspace{-1mm}
\frac{\tau_{l + 1}}{\tau_{l}}
= \frac{\tau_{L^{2}} \sqrt{C_{L^{2}} h_{l}^{d} \#\N{L}}}{\tau_{L^{2}} \sqrt{C_{L^{2}} h_{l + 1}^{d} \#\N{L}}}
= \frac{\sqrt{h_{l}^{d}}}{\sqrt{h_{l + 1}^{d}}}
\simeq \frac{\sqrt{2^{-l d}}}{\sqrt{2^{-(l + 1) d}}}
= \sqrt{2^{d}}
.
\end{equation*}
We denote by \(\kappa\) this scaling factor \(\sqrt{2^{d}}\).

To be consistent with existing works~\cite{sz-reg, zfp, sz-hybrid}, which aim at maximizing PSNR (minimizing \(L^{2}\) error) while respecting an absolute error tolerance (a bound on the \(L^{\infty}\) norm of the error), we next adapt this quantization strategy to control \(L^{\infty}\) error.
It is shown in~\cite{ainsworth2019multilevel} that, for \(C_{L^{\infty}}\) a constant depending on the grid hierarchy,
\vspace{-1mm}
\begin{equation}
\vspace{-1mm}
\begin{split}
&\text{if} \quad \sum_{l = 0}^{L} \max_{x \in \N*{l}} {\abs{\uMC\at{x} - \newMC\at{x}}} \leq \frac{\tau_{L^{\infty}}}{C_{L^{\infty}}}
,\\
&\text{then} \quad \norm{u - \rr{u}}_{L^{\infty}} \leq \tau_{L^{\infty}}
.
\end{split}
\label{eqn:LInfNormEstimator}
\end{equation}
The optimal quantization bin widths in this scenario, found by taking this error condition as a constraint and minimizing \(\cost(q_{0}, \dotsc, q_{L})\) as before, are \(q_{l} = 2 \tau_{L^{\infty}} / C_{L^{\infty}} \times \#\N*{l} / \#\N{L}\).
We note these values only for completeness; in our implementation we will use the geometric scaling \(\tau_{l} = \kappa^{l} \tau_{0}\) obtained in the \(L^{2}\) case.
We next choose \(\tau_{0}\) so that the \(L^{\infty}\) error is bounded by \(\tau_{L^{\infty}}\).
We have
\vspace{-1mm}
\begin{align*}
\vspace{-1mm}
\mathrlap{%
\sum_{l = 0}^{L} \max_{x \in \N*{l}} {\abs{\uMC\at{x} - \newMC\at{x}}}%
}\hphantom{%
\sum_{l = 0}^{L} \kappa^{l} \tau_{0}
}\\
&\leq \sum_{l = 0}^{L} \tau_{l}
= \sum_{l = 0}^{L} \kappa^{l} \tau_{0}
= \frac{1 - \kappa^{L + 1}}{1 - \kappa} \tau_{0}
.
\end{align*}
The righthand term is equal to \(\tau_{L^{\infty}} / C_{L^{\infty}}\) when \(\tau_{0} = (1 - \kappa) / (1 - \kappa^{L + 1}) \times \tau_{L^{\infty}} / C_{L^{\infty}}\).
So, using Eq.~(\ref{eqn:LInfNormEstimator}), if we set the level \(l\) quantization error tolerance to
\begin{equation*}
\tau_{l} = \frac{(1 - \kappa) \kappa^{l}}{1 - \kappa^{L + 1}} \frac{\tau_{L^{\infty}}}{C_{L^{\infty}}}
\quad \text{for} \quad
l = 0, 1, \dotsc, L
,
\label{eq:LinfTau}
\end{equation*}
then \(\norm{u - \rr{u}}_{L^{\infty}} \leq \tau_{L^{\infty}}\), as desired. 
In the rest of the paper, we use $\tau$ for $\tau_{L^{\infty}}$ for simplicity unless specifically noted.

\subsection{Adaptive decomposition}\label{subsec:decomposition}
We further propose to perform multilevel decomposition in an adaptive fashion, i.e., terminating the decomposition procedure at an appropriate level instead of decomposing all the way to the end. 
The reason is two-fold. On the one hand, we note that the piecewise multilinear interpolation is not always better than existing prediction methods such as the Lorenzo predictor used in SZ~\cite{sz17}, especially when the user-specified error tolerance is relatively low. 
On the other hand, the tolerance for the level coefficients is supposed to exponentially decay as the level increases, as we derived in the last subsection. 

As for the remaining coarse-grained representation, we propose to leverage existing error-bounded compression methods~\cite{sz-hybrid, sz17, sz-reg} to deal with them, so that the multilevel approach is used as a preconditioner instead of a standalone compressor. 
In particular, we select SZ~\cite{sz-reg} as our external compressor, because
\begin{enumerate*}
\item it is one of the state-of-the-art error-bounded lossy compressors, leading to high compression quality;
\item it yields the best compression ratio given a fixed error tolerance, according to existing studies~\cite{sz17}; and
\item its Lorenzo predictor is complementary to the piecewise multilinear interpolation used in the multilevel decomposition in terms of prediction efficiency, as will be detailed later in this subsection.
\end{enumerate*}

We compare the effectiveness of the Lorenzo predictor and piecewise multilinear interpolation to determine the appropriate level at which to terminate the multilevel decomposition, because they are the most critical components in the SZ compressor and multilevel compressor, respectively.
In the following text, we first discuss the pros and cons of the Lorenzo predictor used in SZ and the piecewise multilinear interpolation used in the multilevel decomposition. Note that we do not discuss the regression-based predictor used in the current SZ implementation~\cite{sz-reg}, as we observed its prediction accuracy to usually be lower than that of piecewise multilinear interpolation. 
We will then introduce our error estimation method and the adaptive mechanism to automatically terminate the multilevel decomposition.

\subsubsection{Lorenzo versus piecewise multilinear interpolation}
\begin{figure}[ht] \centering
\vspace{-4mm}
\hspace{-4mm}
\subfigure[{Lorenzo predictor}]
{
\includegraphics[width=0.4\columnwidth]{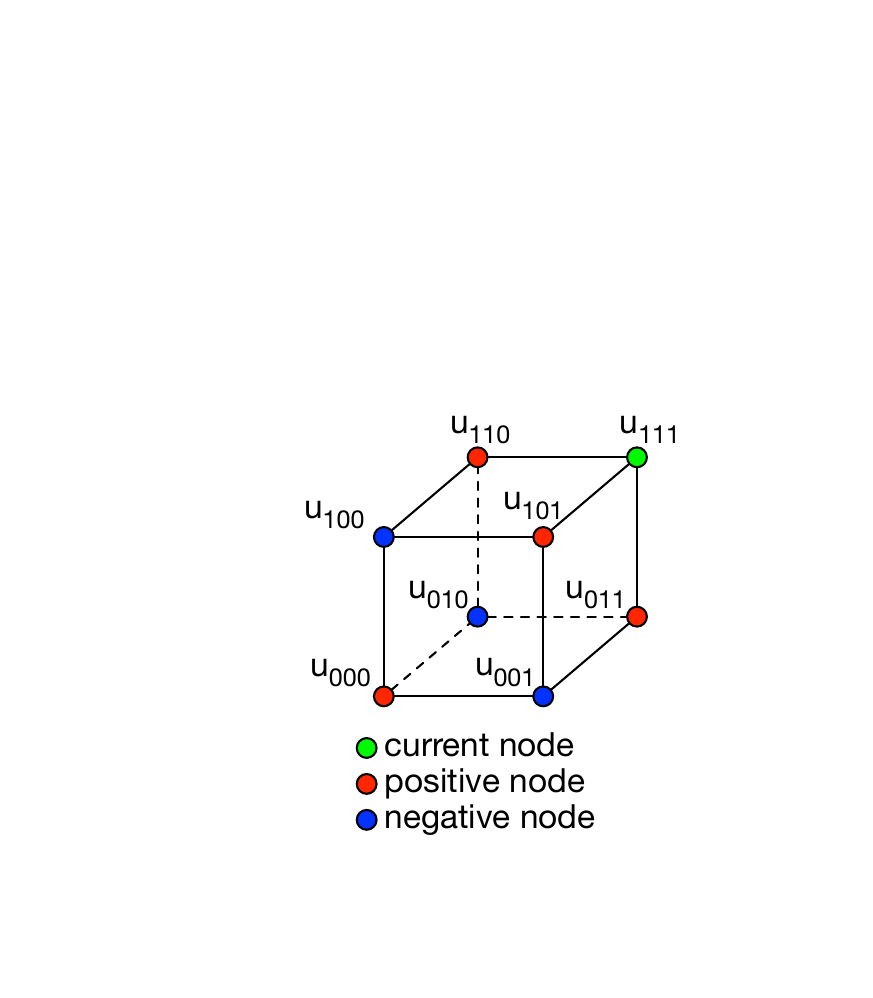}
}
\hspace{-4mm}
\subfigure[{Trilinear interpolation}]
{
\includegraphics[width=0.6\columnwidth]{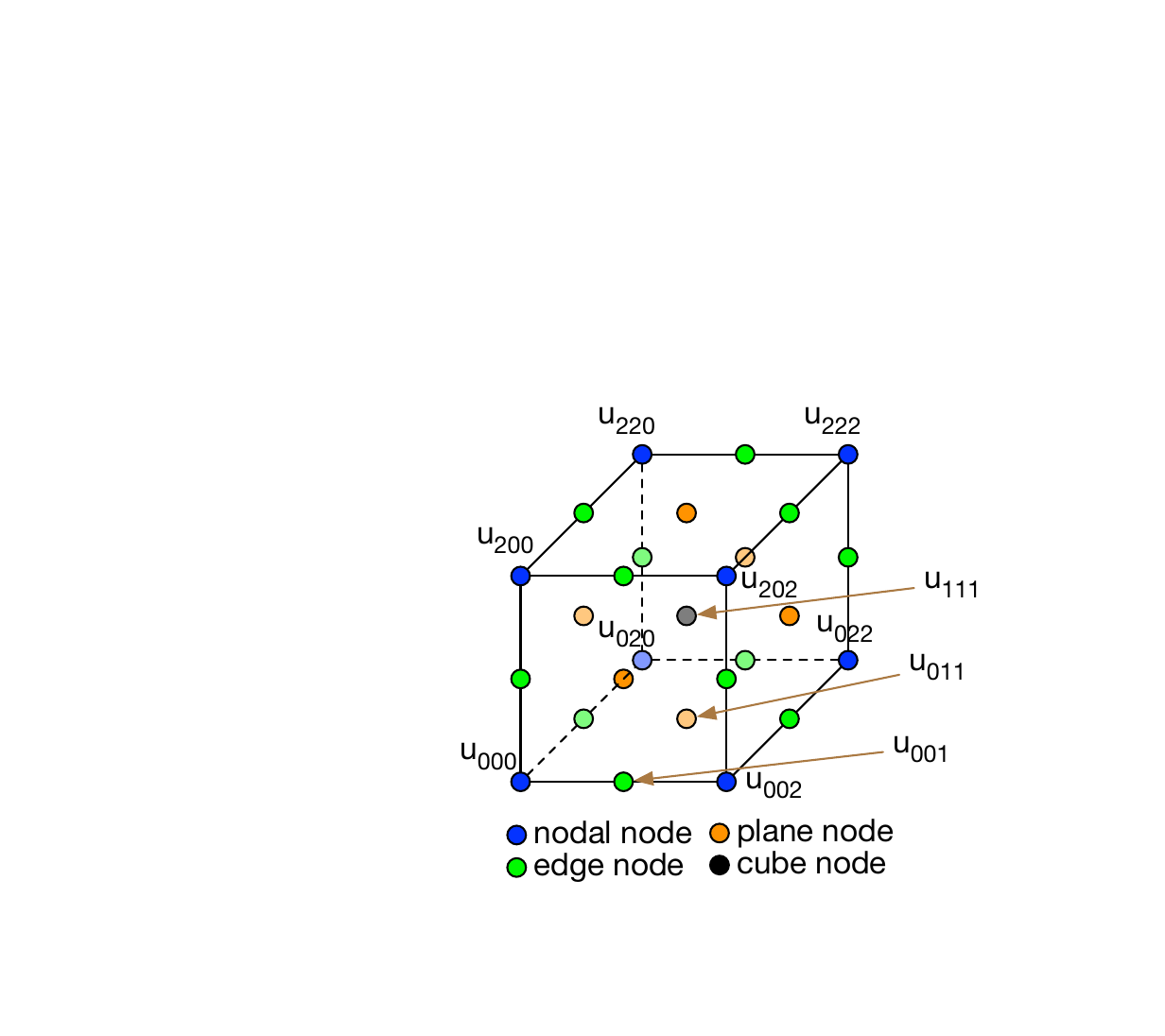}
}
\hspace{-4mm}
\caption{Illustration of Lorenzo predictor and trilinear interpolation.}
\label{fig:prediction}
\end{figure}

The Lorenzo predictor~\cite{lorenzo} estimates the value for a given data point using its immediate neighbors that have already been processed. 
The data value is predicted by the $d-1$ degree polynomial determined by $2^d-1$ neighboring values for $d$-dimensional data, and it is proved that the predicted value can be represented as a signed combination of the neighboring values. 
Thus, the neighboring nodes can be divided into two groups: positive nodes, with sign coefficient $1$, and negative nodes, with sign coefficient $-1$.
For example, as shown in Fig.~\ref{fig:prediction}(a), the 3D Lorenzo predictor predicts the value for current node $u_{111}$ with the following formula:
\begin{equation*}
    \text{pred}_{111}^{\text{Lorenzo}} = u_{110} + u_{101} + u_{011} - u_{100} - u_{010} - u_{001} + u_{000}. 
\end{equation*}
One characteristic of the Lorenzo predictor is that its prediction accuracy relies heavily on the user-specified error tolerance.
Specifically, it has to use reconstructed data with certain errors (i.e., $\Tilde{u}$ instead of $u$) in its prediction, in order to ensure that the predicted values are exactly the same between the compression and decompression stages. This inevitably results in inaccurate predictions, especially when the user-specified error tolerance is relatively high.
On the other hand, the Lorenzo predictor features a high-order data approximation (e.g., using quadratic polynomials for 3D data), which offers accurate prediction, especially when the user-specified error tolerance is low~\cite{lorenzo, sz17}, since the impact of using reconstructed rather than original values is tiny in this situation.

In comparison with the Lorenzo predictor, the piecewise multilinear interpolation is relatively insensitive to errors in the reconstructed data, but has lower prediction accuracy due to the low order of its approximation function.
Specifically, the piecewise multilinear interpolation approximates the value of a data point linearly using the nodal nodes which are present in the next-level subgrid. 
Because the multilevel decomposition uniformly selects half of the nodes in the current level along each dimension as the nodal nodes, it yields independent prediction for each $3^n$ grid. 
For the 3D case illustrated in Fig.~\ref{fig:prediction}(b), the 19 coefficient codes within the $3\times3\times3$ grid can be classified into 3 main categories: edge nodes, which are located on the edge connecting two nodal nodes (e.g., $u_{001}$); plane nodes, which are located in the middle of four nodal nodes (e.g., $u_{011}$); and cube nodes, which are located in center of eight nodal nodes (e.g., $u_{111}$).
The prediction formulas for one example of each category are as follows:
\vspace{-2mm}
\begin{align}\label{eq:multilinear_interp}
\vspace{-2mm}
    \text{pred}^{\text{interp}}_{001} &= \frac{1}{2}(u_{000} + u_{002})\nonumber\\
    \text{pred}^{\text{interp}}_{011} &= \frac{1}{4}(u_{000} + u_{002} + u_{020} + u_{022})\\
\begin{split}
    \text{pred}^{\text{interp}}_{111} &= \frac{1}{8}(u_{000} + u_{002} + u_{020} + u_{022}\nonumber \\
    &\quad\qquad {} + u_{200} + u_{202} + u_{220} + u_{222})
    .
\end{split}
\end{align}

Such characteristics of the two prediction methods inspire us to select the better in between while accounting for the impact of reconstructed data, which determines the most appropriate level to terminate the multilevel decomposition.

\subsubsection{Penalty estimation}
We use penalty factors to efficiently compare the effectiveness of different prediction methods without computing the reconstructed values. 
The penalty factor is defined as the expected difference between the prediction made using the original data and that made using the reconstructed data.
It indicates the degree to which the prediction accuracy will be affected by the user-specified error tolerance.

We briefly introduce the penalty factor for the Lorenzo predictor as follows, as it has been introduced in \cite{sz-reg}. 
Denote $\tau$ as the required error tolerance. Assuming a uniform distribution $U(-\tau, \tau)$ for the errors of decompressed data (which is usually true due to the linear quantization~\cite{sz17}), the penalty factor can be computed via Monte-Carlo method. 
According to~\cite{sz-reg}, the 3D Lorenzo predictor yields a penalty factor of $1.22\tau$. 
Thus, the prediction error of the 3D Lorenzo predictor can be estimated from the original data and the penalty factor as:
\begin{align}
\begin{split}
    E_{\text{Lorenzo}} &= \lvert (u_{110} + u_{101} + u_{011} - u_{100} - u_{010} - u_{001}\\ 
    &\qquad {} + u_{000}) - u_{111}\rvert + 1.22\tau
.
\end{split}
\label{eq:lorenzo}
\end{align}

In the following, we propose the penalty factor for piecewise multilinear interpolation. 
Following the previous approach, we assume a uniform distribution of errors $U(-\tau, \tau)$ for all the nodes in the current level, such that we can compare the two prediction methods under the same compressibility. 
Let $P_{\text{edge}}$, $P_{\text{plane}}$, and $P_{\text{cube}}$ be the penalty terms for the coefficient nodes belonging to edge, plane, and cube nodes, respectively.
According to the prediction formulas in Eq.~\eqref{eq:multilinear_interp}, the penalty terms for multilinear interpolation turn out to be 
\vspace{-1.5mm}
\begin{equation*}
\vspace{-1.5mm}
    P_{\text{edge}} = \sum_{i=0}^1 \frac{\epsilon_i}{2}, \quad
    P_{\text{plane}} = \sum_{i=0}^3 \frac{\epsilon_i}{4}, \quad \text{and} \quad
    P_{\text{cube}} = \sum_{i=0}^7 \frac{\epsilon_i}{8},
\end{equation*}
where $\{\epsilon_i\}$ are the random variables indicating the errors of the nodal nodes.

The errors of nodal nodes in the multilinear decomposition consist of both quantization errors of these nodes and correction errors that are incurred by the quantization errors of coefficient nodes in the current level. 
Using the statistical method, we find that the correction errors for 3D data can be approximated by a Gaussian distribution with mean $0$ and standard deviation $0.283\tau$ when errors of the coefficient nodes are drawn from $U(-\tau, \tau)$, and they are independent from the number of nodes along each dimension. 
These correction errors are then added to the quantization errors on the nodal nodes, which are also drawn from $U(-\tau, \tau)$, to model the penalty. 
Based on our experiments, the penalty factors for coefficient nodes in the three categories on 3D datasets are $E(|P_{\text{edge}}|) = 0.369\tau$, $E(|P_{\text{plane}}|) = 0.259\tau$, and $E(|P_{\text{cube}}|) = 0.182\tau$.
Accordingly, the prediction error of the multilinear approach for the cube node $u_{111}$ in Fig.~\ref{fig:prediction}(b), for instance, can be estimated as:
\vspace{-1.5mm}
\begin{align}
\vspace{-1.5mm}
\begin{split}
    E_{\text{interp}} &= \big\lvert \frac{1}{8}(u_{000} + u_{002} + u_{020} + u_{022} + u_{200} + u_{202}\\
    &\qquad {} + u_{220} + u_{222}) - u_{111}\big\rvert + 0.182\tau
.
\end{split}
\label{eq:interp}
\end{align}

With the computed penalty factors, we can use the original data to estimate the prediction accuracy of the two methods. 
We can also infer that the Lorenzo predictor will be affected more than piecewise multilinear interpolation by using reconstructed data because it has a larger penalty factor, which is consistent with our observation that Lorenzo predictor suffers more on relatively high error tolerance. 

\subsubsection{Adaptive decomposition based on error estimation}
We then use a sampling approach to determine the most appropriate decomposition level based on the estimated accuracy of the two prediction methods. In particular, we uniformly sample the data at the current level in the granularity of $3^d$ blocks and estimate the prediction errors of coefficient nodes for both the Lorenzo predictor and piecewise multilinear interpolation.
Note that we need to apply the penalty factors computed above to account for the impact of reconstructed data.
In our implementation, we sample one out of four blocks along each dimension and aggregate the estimated prediction errors for each of the two prediction methods. 
If the aggregated prediction error of the Lorenzo predictor is lower than that of piecewise multilinear interpolation, we compress all of the data in the current level using SZ (which uses the Lorenzo predictor) and terminate the multilevel decomposition; otherwise, we will continue to perform the multilevel decomposition.

\begin{algorithm}[ht]
\caption{\textsc{Multilevel data reduction with level-wise quantization and adaptive decomposition}} \label{alg:compression} \footnotesize
\renewcommand{\algorithmiccomment}[1]{/*#1*/}
\textbf{Input}: $d$-dimensional data $u$, required global error tolerance $\tau$\\
\textbf{Output}: compressed data $s$

\begin{algorithmic} [1]
\STATE $Q_L \gets I$, $\Tilde{l} \gets 0$, $\kappa \gets \sqrt{2^d}$
\FOR{$l = L \to 0$}
	\STATE $\tau_0 \gets \frac{(1-\kappa)\tau}{(1-\kappa^{L+1-l})C_{L^{\infty}}}$
	\STATE $\Hat{u} \gets \texttt{block\_based\_sampling}(Q_lu)$
	\STATE $E_{\text{Lorenzo}} \gets 0, E_{\text{interp}} \gets 0$
	\FOR {$x \in {\N*{l}} \cap \Hat{u}$}
	    \STATE $E_{\text{Lorenzo}} \gets E_{\text{Lorenzo}} + \texttt{estimate\_Lorenzo\_err}(x, \tau_0)$
	    \STATE $E_{\text{interp}} \gets E_{\text{interp}} + \texttt{estimate\_interp\_err}(x, \tau_0)$
	\ENDFOR
	\IF{$E_{\text{Lorenzo}} < E_{\text{interp}}$}
	    \STATE $\Tilde{l} = l$
	    \STATE \textbf{break}
	\ELSE
	    \STATE $Q_{l-1}u, \uMC\at{\N*{l}} \gets \texttt{multi\_grid\_decomposition}(Q_lu)$
    \ENDIF
\ENDFOR
\STATE $\tau_{\Tilde{l}} \gets \frac{(1-\kappa)\tau}{(1-\kappa^{L+1-\Tilde{l}})C_{L^{\infty}}}$
\STATE $s_0 \gets \texttt{external\_lossy\_compress}(Q_{\Tilde{l}}u, \tau_{\Tilde{l}})$
\FOR{$l = \Tilde{l} + 1 \to L$}
    \STATE $\tau_l \gets \kappa\tau_{l-1}$
    \STATE $\newMC\at{\N*{l}} \gets \texttt{quantize}(\uMC\at{\N*{l}}, \tau_l)$
\ENDFOR
\STATE $s_1 \gets \texttt{lossless\_compress}(\newMC)$
\STATE $s \gets \texttt{concat}(s_0, s_1)$
\RETURN $s$
\end{algorithmic}
\end{algorithm}

Algorithm~\ref{alg:compression} presents the pseudo-code of our proposed compression algorithm, which is described as follows. 
After initializing the necessary variables (line~1), we perform the adaptive multilevel decomposition level by level (lines~2$\sim$16). 
Before performing the decomposition, we compute the theoretical error tolerance $\tau_0$ and perform block-based sampling for data in the current level (lines~3 and~4, respectively). 
Then, we iterate through all the coefficient nodes in the sampled data (line~6) and accumulate the prediction errors of the Lorenzo predictor (line~7) and multilinear interpolation (line~8), with the estimation functions in Eq.~\eqref{eq:lorenzo} and Eq.~\eqref{eq:interp}. 
If the prediction error of the Lorenzo predictor is less than that of multilinear interpolation, we will stop the decomposition and switch to an external compressor (SZ in this case) to compress the remaining coarse representation. 
Otherwise, we will decompose data with the multilevel method~\cite{ainsworth2019multilevel} and move on to the next level. 

After the multilevel decomposition terminates, we perform the level-wise quantization (lines~17$\sim$23). 
Specifically, we derive the error tolerance for the coarse-grained representation and compress it with an external compressor (lines~17$\sim$18). 
After that, we iterate through all the decomposition levels, multiply the current error tolerance with the scaling factor and use the updated error tolerance to quantize multilevel coefficients $\uMC$ in each level. 
Finally, the quantized multilevel coefficients $\newMC$ are compressed losslessly and concatenated with the compressed coarse-grained representation to generate the compressed format.
\vspace{-5mm}
\section{Implementation and Optimizations}\label{sec:optimization}
Besides quality, performance is another important aspect for large-scale scientific data compression.
In this section, we introduce a series of optimizations that we adopt to improve the performance and efficiency of multilevel data decomposition/recomposition algorithms. 

\subsection{Level-centric data reordering}\label{subsec:reorder}
We leverage a data reordering algorithm to deal with the strided memory access in the multilevel method, inspired by the de-interleaving phase in wavelet decomposition.
Specifically, we rearrange the input data in a level-centric manner to put the nodal nodes in coherent memory space, so that the memory accesses in the next level can be efficient.
In what follows, we first identify the data dependencies and memory access patterns in the multilevel decomposition and introduce the reordering algorithm thereafter.

We illustrate the data dependencies of the key steps (coefficient computation, correction computation, and correction application) of the multilevel decomposition in Fig.~\ref{fig:dependency} using a 2D example.
In each iteration, \textit{coefficient computation} computes the piecewise multilinear interpolations for all the coefficient nodes using their adjacent nodal nodes in the current level, and updates the values of these coefficient nodes with the difference between their original values and the piecewise multilinear interpolations. 
After that, \textit{correction computation} performs a row sweep to compute the row correction, followed by a column sweep on the resulting row correction to obtain the overall correction. 
At last, the overall correction is applied to the nodal values in the \textit{correction application}. 
As we can see from the figure, almost all the operations here involve strided memory access which skips the processed nodes, leading to inefficient cache utilization because values of processed nodes are fetched but never accessed. 
In other words, only the first level of the decomposition makes efficient use of the cache.
Furthermore, a stride larger than the number of elements that a cacheline can hold will result in cache misses everywhere.

\begin{figure}[ht] \centering
\vspace{-3mm}
\includegraphics[width=\columnwidth]{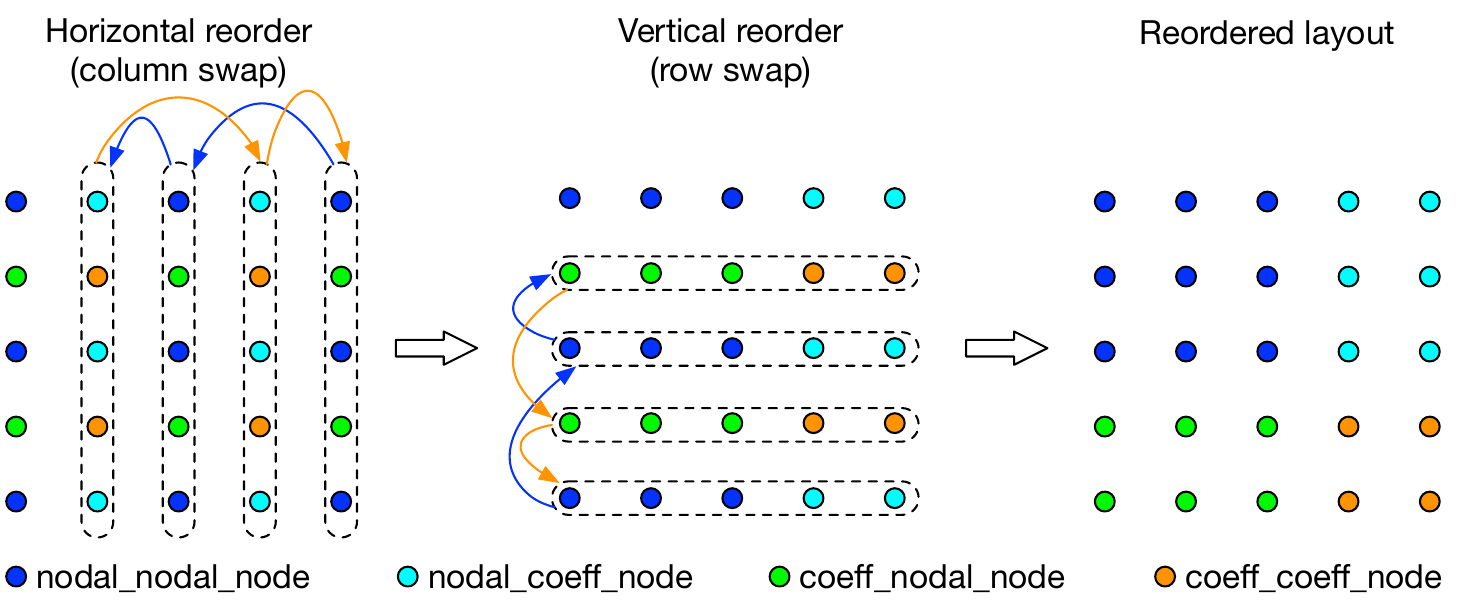}
\vspace{-6mm}
\caption{2D data reordering.}
\label{fig:reorder_example_2d}
\vspace{-2mm}
\end{figure}

\newcommand\nodalnodalnode{\texttt{nodal\_nodal\_node}}
\newcommand\nodalcoeffnode{\texttt{nodal\_coeff\_node}}
\newcommand\coeffnodalnode{\texttt{coeff\_nodal\_node}}
\newcommand\coeffcoeffnode{\texttt{coeff\_coeff\_node}}

To enable efficient cache utilization, we reorder the data in order to cluster the nodal nodes (i.e., the nodes that will be used for decomposition at the next level) in the current level together. 
We implement this by a horizontal reordering and a vertical reordering. 
For a 2D grid of $(2n_1 + 1) \times (2n_2 + 1)$, the horizontal reordering essentially moves the $(2i + 1)$-th column to the $i$-th column and the $2i$-th column to the $(2i + 1)$-th column, where $i$ ranges from $1$ to $n_2$, such that the nodal columns are moved together. 
Then, the vertical reordering applies similar operations to the rows in order to cluster the nodal rows together. 
Denote \nodalnodalnode\ as nodes in both nodal rows (i.e., rows that will be present in the next-level subgrid) and nodal columns (i.e., columns that will be present in the next-level subgrid), \nodalcoeffnode\ as nodes in nodal rows but coefficient columns (i.e., columns that will be absent in the next-level subgrid), \coeffnodalnode\ as nodes in coefficient rows (i.e., rows that will be absent in the next-level subgrid) but nodal columns, and \coeffcoeffnode\ as nodes in both coefficient rows and coefficient columns. Fig.~\ref{fig:reorder_example_2d} shows how the reordering algorithm works for $5\times5$ 2D data.

\begin{figure}[ht] \centering
\includegraphics[width=\columnwidth]{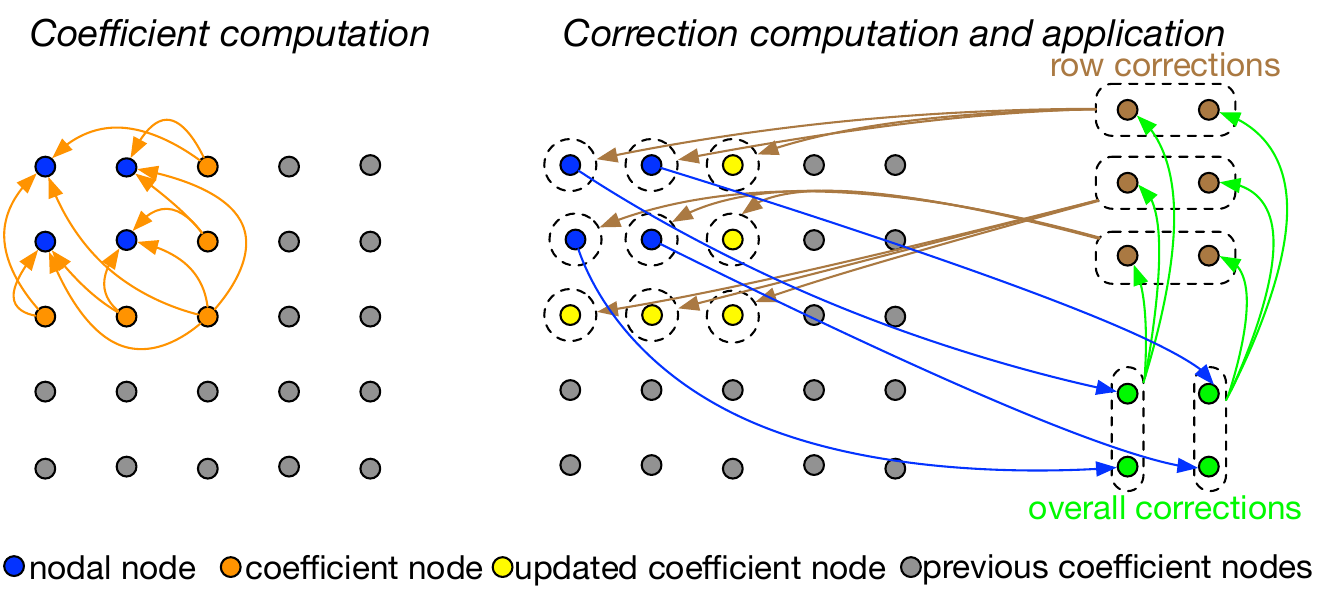}
\vspace{-4mm}
\caption{Reordered data layout and dependencies ($l = L-1$).}
\label{fig:reordered_dependency}
\vspace{-2mm}
\end{figure}

Fig.~\ref{fig:reordered_dependency} shows the data dependencies in the multilevel method after reordering. Compared with the original data, the reordered data requests memory access in a more coherent way, which promises higher performance.

Given the reordered data, we then perform sliding window update for efficient coefficient computation. Specifically, we locate the starting positions of different groups of nodes (e.g., \nodalnodalnode, \nodalcoeffnode, \coeffnodalnode, and \coeffcoeffnode\ for the 2D case), and compute their coefficients simultaneously. In this way, coefficient computation can be performed more efficiently with relatively high cache utilization.

During recomposition, data is already ordered in a level-centric manner. In this case, we perform correction computation, inverse correction application, and inverse coefficient computation on the ordered data directly, followed by an inverse reordering operation to put recomposed data to the correct positions of the finer level.

\subsection{Direct load vector computation} \label{subsec:load_vector}
\begin{figure}[ht] \centering
\vspace{-2mm}
\hspace{-4mm}
\subfigure[{1D}]
{
\includegraphics[width=0.5\columnwidth]{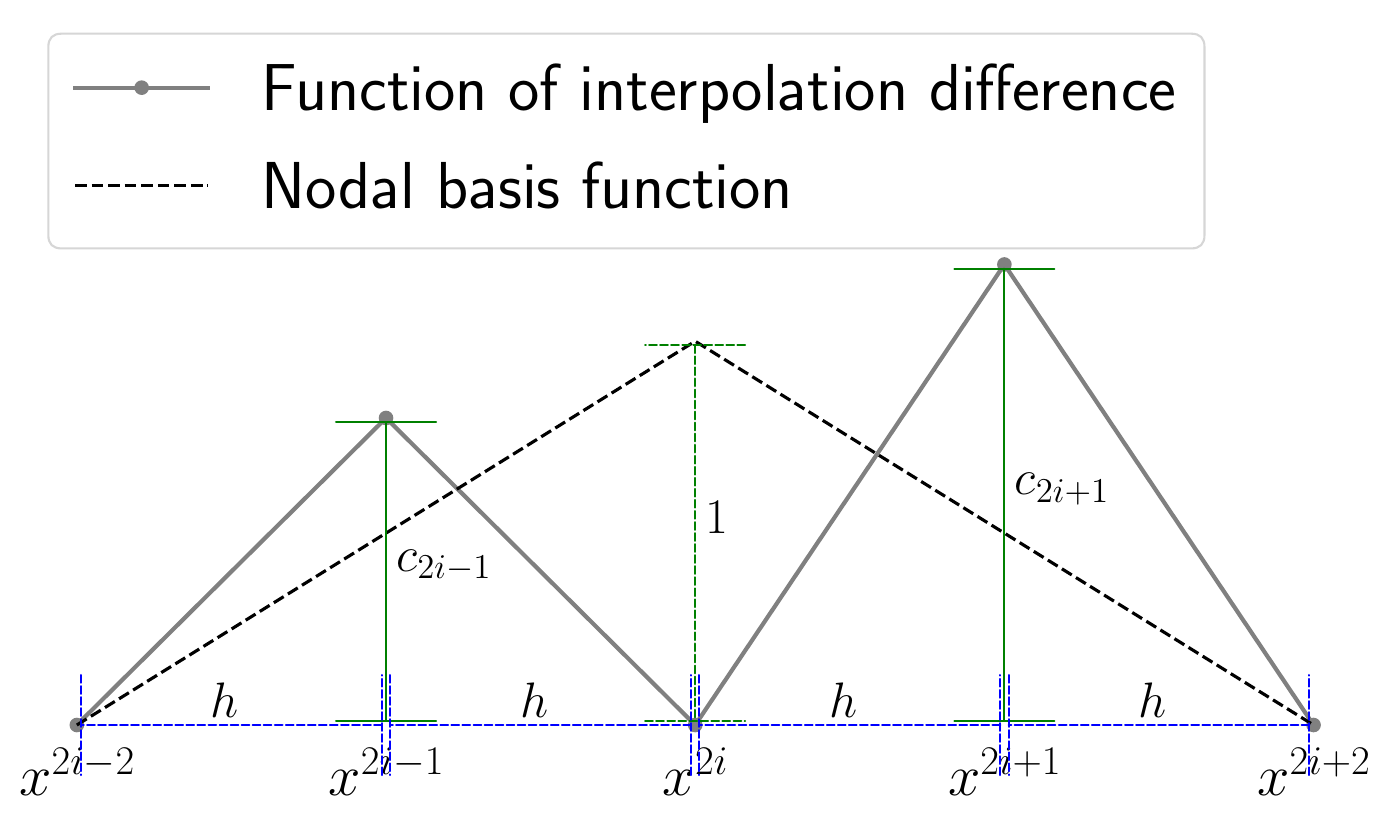}
}
\hspace{-4mm}
\subfigure[{generalized}]
{
\includegraphics[width=0.5\columnwidth]{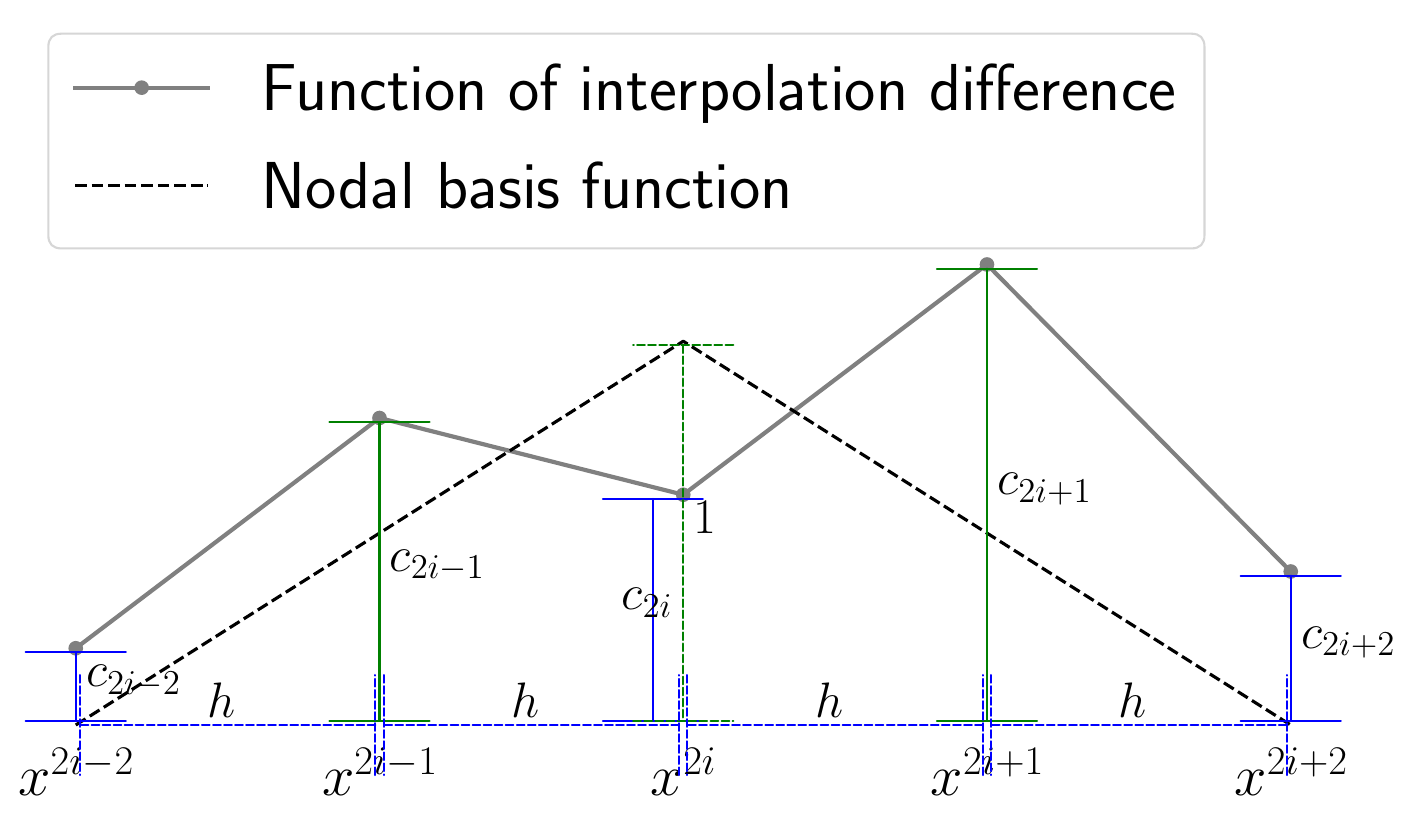}
}
\hspace{-4mm}
\caption{1D and generalized load vector computation.}
\label{fig:load_vector}
\end{figure}

We next derive a formula for load vector computation in the correction computation to reduce computational cost, which was computed by fine-grained mass matrix multiplication followed by a restriction transform in previous multilevel method. 
The load vector is defined as the function expressed in terms of nodal displacements, and it is computed by the inner product of the function representing interpolation difference and the nodal basis function.
Although the multidimensional load vector computation boils down to multiple 1D computations along each dimension, it is not exactly the same as that of 1D cases as displayed in Fig.~\ref{fig:load_vector}.
Specifically, non-zero interpolation differences only appear on the coefficient nodes in 1D case, because the interpolations on nodal nodes are equal to the nodal values. 
However, this is not the case for the multidimensional cases, because the coefficient nodes may be nodal nodes along a certain dimension during the computation (e.g., \nodalcoeffnode\ in Fig.~\ref{fig:reorder_example_2d}).
To tackle this issue, we generalize the direct load vector computation with the following lemma.
\begin{lemma}\label{lem:dlvc}
Denote the values of nodes in the current level as $c_0, c_1, \dots, c_{2n+1}$ as shown in Fig.~\ref{fig:load_vector}. Further denote $c_{-1} = c_{2n+2} = 0$ and the internode spacing in the current level as $h_l$.
The $i$-th entry of the load vector in generalized case can be computed by
\vspace{-2mm}
\begin{equation*}\label{eq:load_vector}
    f_i = (\frac{1}{12}c_{2i-2} + \frac{1}{2}c_{2i-1} + \frac{5}{6}c_{2i} + \frac{1}{2}c_{2i+1} + \frac{1}{12}c_{2i+2})h_l.
\vspace{-2mm}
\end{equation*}
\end{lemma}
We omit the proof as it is derived via direct computation of the corresponding integrals. 
When the nodal values $\{c_{2i}\}$ are all $0$, the lemma degrades to the 1D case derived in~\cite{ainsworth2018multilevel}.

\subsection{Batched correction computation}\label{subsec:batch}
We use batch operations to further improve the memory access efficiency for the intermediate corrections, in addition to that for the node values optimized in Section~\ref{subsec:reorder}. 
As indicated by the green arrows in Fig.~\ref{fig:dependency} and Fig.~\ref{fig:reordered_dependency}, the column sweep that computes the load vector and solves the corresponding tridiagonal linear system for each separate column of the row corrections requires strided memory access.
This problem is exacerbated on 3D datasets, which have such memory access patterns along two discontiguous dimensions. 
Fortunately, the column sweep needs to be applied to all the columns in the current level and thus can be optimized through batch operations. 
Specifically, denoting by \(b\) the batchsize, we perform direct load vector computation on $b$ contiguous nodes simultaneously. 
In other words, the $i$-th entries of the load vectors for $b$ columns are computed together to achieve high cache efficiency. After that, a slightly modified Thomas algorithm is used to solve the tridiagonal linear systems in batch. 

This optimization requires $b\max{n_i} / \prod_i{n_i}$ extra memory to store the column load vectors, where $n_i$ is the number of data points along the $i$-th dimension. Because the batchsize $b$ is usually small, such memory overhead is indeed negligible for multi-dimensional cases.
In our experiments, we observe that $b = 16$ already yields good performance that is similar to that of any larger batch sizes. To account for variations on different systems while limiting
memory overhead, we use $b = 32$ as our default configuration.

\subsection{Intermediate variable elimination \& reuse}\label{subsec:other_opt}
We identify and eliminate the common multipliers as well as repeat computation of intermediate variables in the multilevel algorithm for more efficient computation. 

First, we found that the internode spacing $h_{l}$ is the common multiplier in the tridiagonal linear system solving for the correction computation. Specifically, the target linear system can be written as follows:
\begin{equation*}
\resizebox{\linewidth}{!}{%
\small
    $\begin{bmatrix}
        \frac{2}{3}{\color{blue}h_{l}} & \frac{1}{3}{\color{blue}h_{l}} & 0 & \hdots & 0 & 0 & 0\\
        \frac{1}{3}{\color{blue}h_{l}} & \frac{4}{3}{\color{blue}h_{l}} & \frac{1}{3}{\color{blue}h_{l}} & \hdots & 0 & 0 & 0\\
        0 & \frac{1}{3}{\color{blue}h_{l}} & \frac{4}{3}{\color{blue}h_{l}} & \hdots & 0 & 0 & 0\\
        \vdots & \vdots & \vdots & \ddots & \vdots & \vdots & \vdots\\
        0 & 0 & 0 & \hdots & \frac{1}{3}{\color{blue}h_{l}} & \frac{4}{3}{\color{blue}h_{l}} & \frac{1}{3}{\color{blue}h_{l}} \\
        0 & 0 & 0 & \hdots & 0 & \frac{1}{3}{\color{blue}h_{l}} & \frac{2}{3}{\color{blue}h_{l}}\\
    \end{bmatrix}  
    \begin{bmatrix}
    x_0\\
    x_1\\
    x_2\\
    \vdots \\
    x_{n-2}\\
    x_{n-1}\\
    \end{bmatrix}
    =
    \begin{bmatrix}
    F_0{\color{blue}h_{l}} \\
    F_1{\color{blue}h_{l}} \\
    F_2{\color{blue}h_{l}} \\
    \vdots \\
    F_{n-2}{\color{blue}h_{l}} \\
    F_{n-1}{\color{blue}h_{l}} \\
    \end{bmatrix},$
}%
\end{equation*}
where $\{F_i = \frac{1}{12}c_{2i-2} + \frac{1}{2}c_{2i-1} + \frac{5}{6}c_{2i} + \frac{1}{2}c_{2i+1} + \frac{1}{12}c_{2i+2}\}$ are the coefficients of the load vector derived in Lemma~\ref{lem:dlvc}, $h_{l}$ is the internode spacing, and $\{x_i\}$ are the corrections that need to be solved.
Thus, the common multiplier $h_{l}$ can be extracted and cancelled from the mass matrix generation and load vector computation to save computational cost.


We also reuse the intermediate variables to avoid repeated computation, where the auxiliary arrays used in solving the tridiagonal linear systems are typical examples. 
Because the tridiagonal mass matrix is fixed for each dimension, we compute the related auxiliary arrays at the very beginning of the multilevel decomposition/recomposition algorithm and pass the precomputed results as parameters. This adjustment reduces the computational complexity on these variables from $O(\prod_{i=0}^d n_i)$ to $O(\sum_{i=0}^d n_i)$, with merely $\sum_{i=0}^d n_i$ additional memory.
\section{Evaluation}\label{sec:evaluation}
In this section, we present the performance evaluation results to demonstrate the effectiveness of our method for scientific data reduction and refactoring. 
Specifically, we compare both compression/decompression performance and rate--distortion of our method (MGARD+) with four state-of-the-art error-bounded lossy compressors --  MGARD~\cite{ainsworth2019multilevel}, SZ~\cite{sz-reg}, ZFP~\cite{zfp}, and the hybrid model proposed in~\cite{sz-hybrid} -- using four real-world datasets from Scientific Data Reduction Benchmarks~\cite{sdrb}. 
We also show how our approach improves the efficiency of scientific analysis using the iso-surface mini-application widely used in scientific visualization. 

\subsection{Experimental Setup}
We conducted our experimental evaluations on the Rhea cluster~\cite{rhea} at Oak Ridge National Laboratory. Each node on the system has two 8-core Intel Xeon E5-2650 processors and 128~GB of memory. 
For all compressors we benchmarked, the latest releases were used as of Sep.~1st, 2020. 
They are compiled with gcc-4.8.5, the default compiler provided in the cluster.
The datasets we use for evaluation are from various domains, including Hurricane Isabel climate simulation~\cite{hurricane}, NYX cosmology simulation~\cite{nyx}, SCALE-LETKF weather simulation~\cite{scale-letkf}, and QMCPACK~\cite{kim2018qmcpack} quantum Monte Carlo simulation. The details of the datasets are listed in Table~\ref{tab:dataset}. 
Note that the data size in the table only accounts for data in a single core.
The total data size goes up to 2.4~TB, 6~TB, 12.6~TB, and 1.2~TB when 2k cores are used in our scalability evaluation.

\begin{table}[ht]
\vspace{-3mm}
\centering
\caption{Datasets for evaluation}
\label{tab:dataset}
\begin{tabular}{|l|c|c|c|c|c|}
\hline
    Dataset & \#Fields & Dimensions & Size\\
\hline
Hurricane Isabel  & 13 & $100\times500\times500$ & 1.21 GB\\
\hline
NYX & 6 & $512\times512\times512$ & 3 GB\\ 
\hline
SCALE-LETKF & 12 & $98\times1200\times1200$ & 6.31 GB\\
\hline
QMCPACK & 1 & $288\times115\times69\times69$ & 0.59 GB\\
\hline
\end{tabular}
\vspace{-3mm}
\end{table}

\subsection{Performance}
We evaluate the performance of our framework in terms of throughput regarding both decomposition/recomposition and compression/decompression. 
Specifically, we first present the performance improvement on the multilevel decomposition/recomposition from the proposed optimizations with detailed breakdown, and how it can be used to accelerate scientific data analytics. 
After that, we compare the overall compression/decompression throughput among the state-of-the-art error-bounded lossy compressors. 
At last, we conduct a parallel experiment to demonstrate the scalability of the algorithm.

\subsubsection{Decomposition/recomposition}
\begin{figure}[ht] \centering
\vspace{-4mm}
\includegraphics[width=0.5\textwidth]{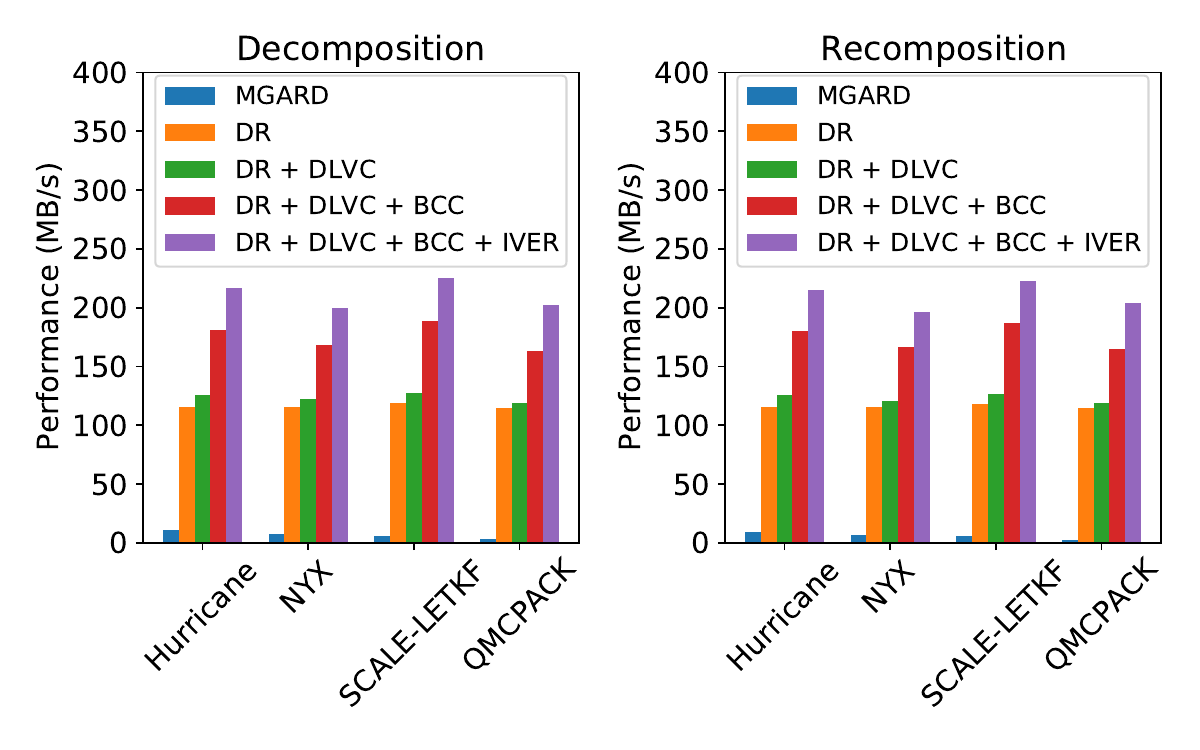}
\vspace{-10mm}
\caption{Decomposition/recomposition performance with optimizations.}
\label{fig:performance}
\end{figure}

Fig.~\ref{fig:performance} illustrates the decomposition/recomposition performance improvement of our framework with respect to the optimizations in Section~\ref{sec:optimization} when compared with existing approach (the blue bar in the figure). 
DR, DLVC, BCC, IVER are the abbreviations for data reordering (Section~\ref{subsec:reorder}), direct load vector computation (Section~\ref{subsec:load_vector}), batched correction computation (Section~\ref{subsec:batch}), and intermediate variable elimination/reuse (Section~\ref{subsec:other_opt}), respectively.
We evaluate the impact of the four optimizations by including them incrementally (corresponding to the orange bar, the green bar, the red bar, and the purple bar in the figure).

From this figure, we observe that the optimizations adopted in our method significantly improve the efficiency of the multilevel method for both decomposition and recomposition. 
In particular, the decomposition performance of our method is $20\times$, $28\times$, $39\times$ and $71\times$ that of the existing multilevel method on the four datasets, respectively. 
Similarly, the recomposition performance of our method is over $22\times$, $30\times$, $41\times$ and $80\times$ that of the existing implementation. 
These performance improvements demonstrate the effectiveness of the proposed optimizations.

\subsubsection{Use case: accelerating iso-surface computation}
We further show how our method benefits scientific analysis by performing analysis on the decomposed representations, using iso-surface computation in scientific visualization as an example. 
An iso-surface is a three-dimensional analog of an iso-line. It is a surface that represents points of a constant value (i.e., iso-value) within a volume of space, which can be used to study specific features around objects, such as features of fluid flows around aircraft wings in computational fluid dynamics. 
In what follows, we first show the accuracy of iso-surface computation using MGARD and our method with different levels of decomposed representations. 
After that, we demonstrate the performance improvement of conducting analysis with those decomposed representations.

We use the area of the iso-surface, which is the outcome of the analysis, to measure the accuracy of different level representations. 
Two representative fields (velocity\_x and temperature) from NYX datasets are evaluated with designated iso-values. 
The iso-value is set to $0$ for velocity\_x because there are specific property scientists would like to see when velocity equals $0$. 
As for temperature, the iso-value is set to the mean of data. 
We perform the multilevel decomposition for $3$ times, which leads to $4$ levels of representations. 
According to our definitions in Section~\ref{sec:background}, level $3$ is the finest-grained representation (original data) while level $0$ is the coarsest-grained one.
Table~\ref{tab:iso_vx} and Table~\ref{tab:iso_temp} display the relative errors on the area of iso-surface for the designated iso-value using the different representations. 
Note that the relative errors of MGARD and our method are not exactly the same because of the different treatments on the non-dyadic cases when data dimensions are not in the form of $2^k+1$ where $k$ is an integer.
To avoid the expensive preprocessing steps for such cases in MGARD, we introduce extra dummy nodes while performing the data reordering, which leads to slightly different decomposition results. 
From the two examples, we can see that our method has similar errors to those of MGARD while offering significantly higher decomposition/recomposition performance. 

\begin{table}[ht]
\vspace{-2mm}
\centering
\caption{Relative error on iso-surface area and decomposition performance (NYX velocity\_x)}
\vspace{-2mm}
\label{tab:iso_vx}
\begin{tabular}{|l|l|c|c|c|c|}
\hline
& Level     & 2 & 1 & 0 \\
\hline
\multirow{2}{*}{Rel. Error} & MGARD & 1.61\% & 0.07\% & 5.23\% \\ 
\cline{2-5}
& MGARD+ & 1.65\% & 0.10\% & 5.21\%\\
\hline
\multirow{2}{*}{Perf. (MB/s)} & MGARD & 8.95  & 7.27 & 6.90 \\
\cline{2-5}
& MGARD+ & 226.63 & 203.72 & 202.67 \\ 
\hline
\end{tabular}
\vspace{-2mm}
\end{table}
\begin{table}[ht]
\vspace{-2mm}
\centering
\caption{Relative error on iso-surface area and decomposition performance (NYX temperature)}
\vspace{-2mm}
\label{tab:iso_temp}
\begin{tabular}{|l|l|c|c|c|c|}
\hline
& Level     & 2 & 1 & 0 \\
\hline
\multirow{2}{*}{Rel. Error} & MGARD & 5.72\% & 7.58\% & 6.86\% \\ 
\cline{2-5}
& MGARD+ & 5.79\% & 7.64\% & 6.89\%\\
\hline
\multirow{2}{*}{Perf. (MB/s)} & MGARD & 8.69 & 7.04 & 6.68 \\
\cline{2-5}
& MGARD+ & 226.39 & 204.60 & 202.66 \\ 
\hline
\end{tabular}
\vspace{-2mm}
\end{table}

\begin{figure}[ht] \centering
\includegraphics[width=0.5\textwidth]{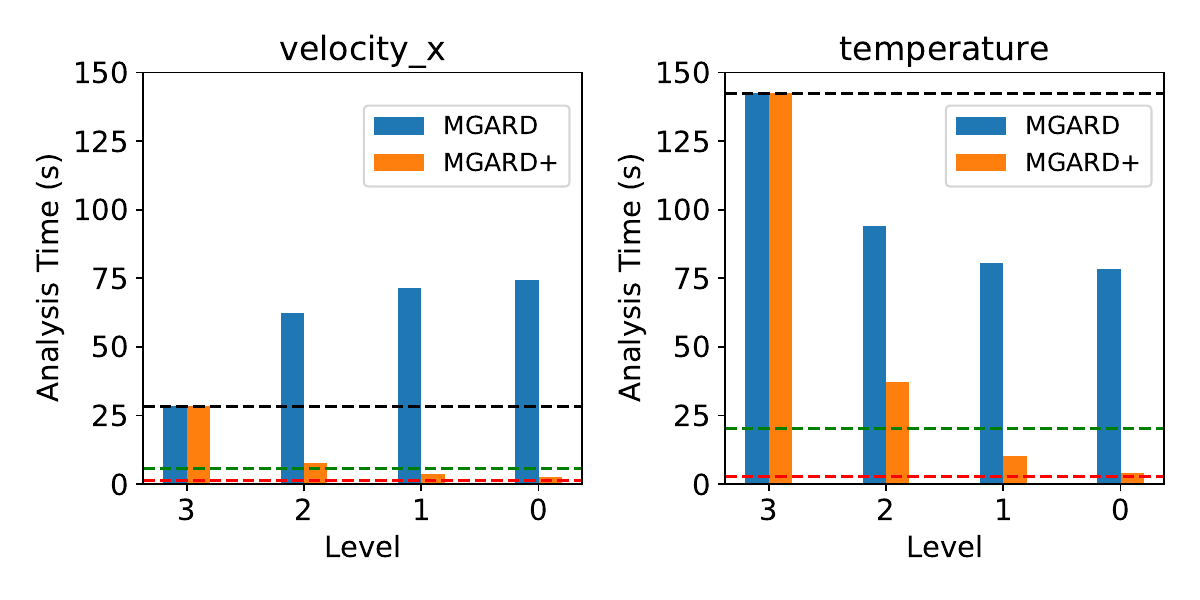}
\vspace{-10mm}
\caption{Overall analysis time (decomposition time plus analysis time on reduced representation) of iso-surface evaluation on two representative fields in the NYX dataset. Dashed lines indicate the strong-scaling result of performing iso-surface analysis with 1 core (black), 8 cores (green) and 64 cores (red).}
\label{fig:analysis_tp}
\vspace{-4mm}
\end{figure}

Fig.~\ref{fig:analysis_tp} displays the overall time spent on analysis, which includes both time for decomposition and time for conducting analysis on the decomposed representation, when using MGARD and our method.
The black dashed line indicates the time for performing the analysis on the original data, and the green and red dashed lines show the time for conducting strong-scaling experiments with 8 cores and 64 cores, respectively. 
It is observed that MGARD suffers from high decomposition overhead, which leads to minor performance gain (temperature) and may even be more costly when the analysis is fast (velocity\_x).
On the other hand, our method significantly improves the performance of the scientific analysis: performing the analysis on level $0$ representation with single core can lead to comparable performance to that of strong-scaling with 64 cores. 
Also, such performance gain would be more significant when multiple iso-values need to be computed (and so the analysis becomes more expensive).

\subsubsection{Compression/decompression}
\begin{figure*}[ht] \centering
\vspace{-1em}
\hspace{-4mm}
\subfigure[{Compression performance}]{
\includegraphics[width=0.5\textwidth]{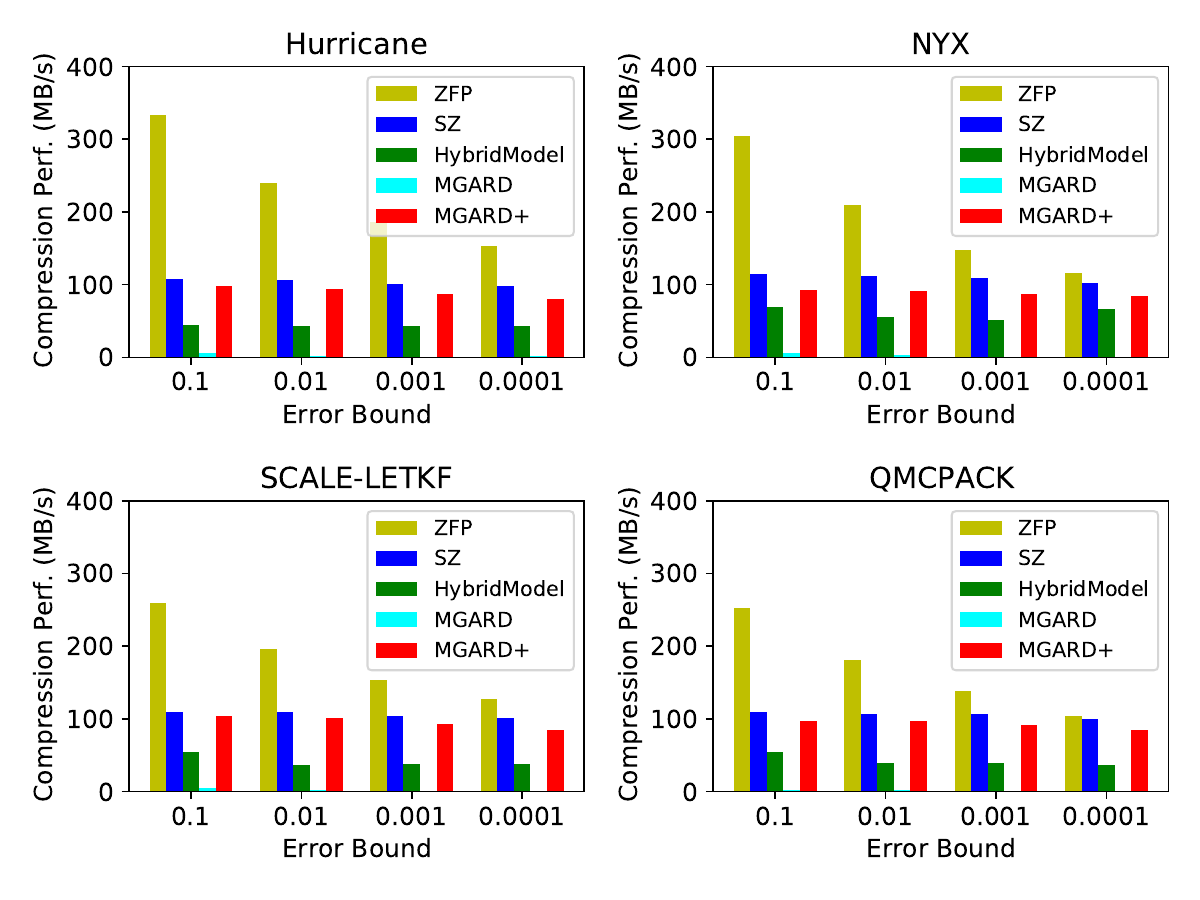}
}
\hspace{-4mm}
\subfigure[{Decompression performance}]{
\includegraphics[width=0.5\textwidth]{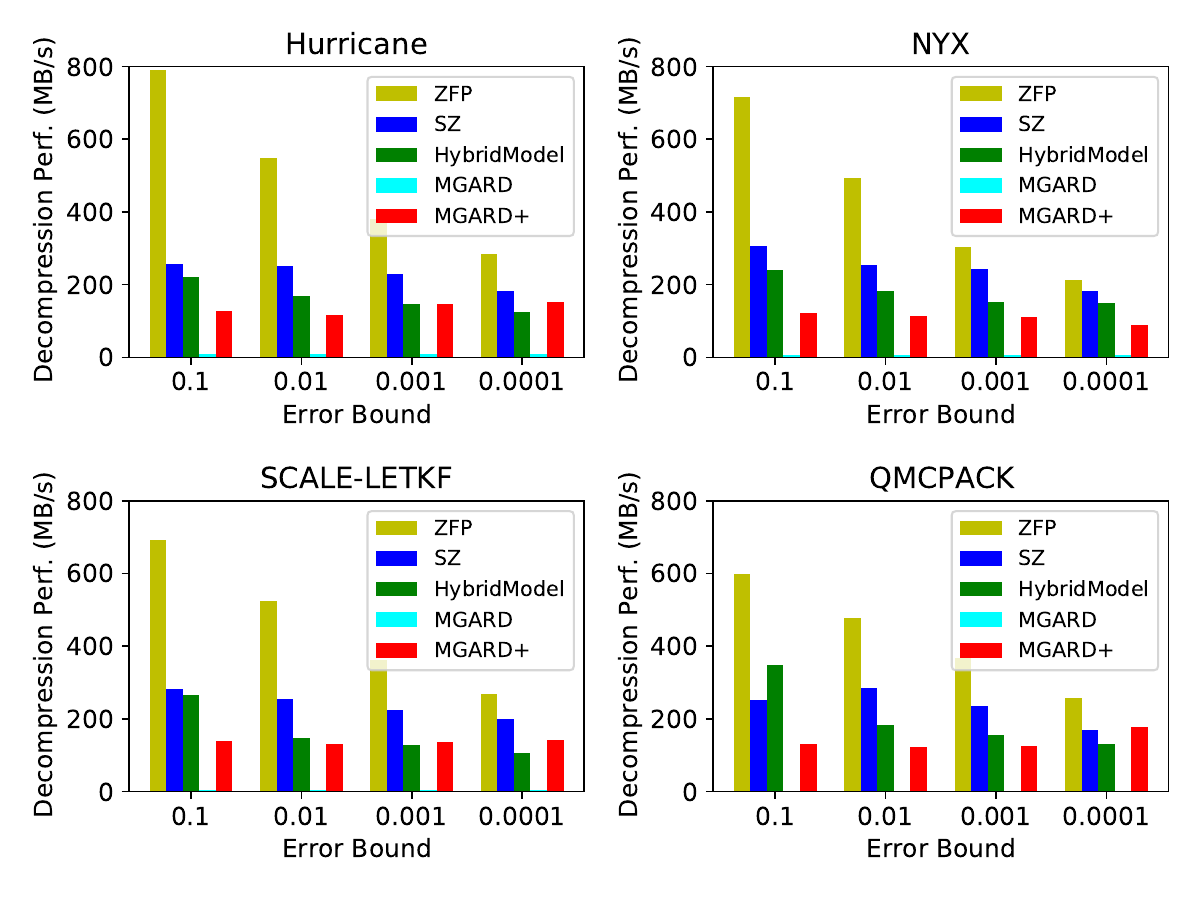}
}
\hspace{-4mm}
\vspace{-2mm}
\caption{Compression/decompression performance of the error-bounded lossy compressors.}
\label{fig:reduction_rate}
\vspace{-1em}
\vspace{-2mm}
\end{figure*}
We then present the compression/decompression performance of our method and compare it with state-of-the-art lossy compressors in Fig.~\ref{fig:reduction_rate}. 
According to this figure, ZFP leads all the evaluated compressors in terms of both compression and decompression performance, but its advantage decreases as the error tolerance becomes low. 
Our method significantly improves the performance of the previous multilevel approach (MGARD), leading to compression performance comparable to that of SZ. 
The decompression performance of our method is lower than SZ, because decompression in our approach is as costly as compression due to the symmetric operations, while SZ has higher decompression performance than compression performance. 
The hybrid model has slightly higher decompression performance than our method, but its compression performance is only one half that of our method in most cases.

\begin{figure}[t] \centering
\vspace{-2mm}
\hspace{-4mm}
\includegraphics[width=0.45\textwidth]{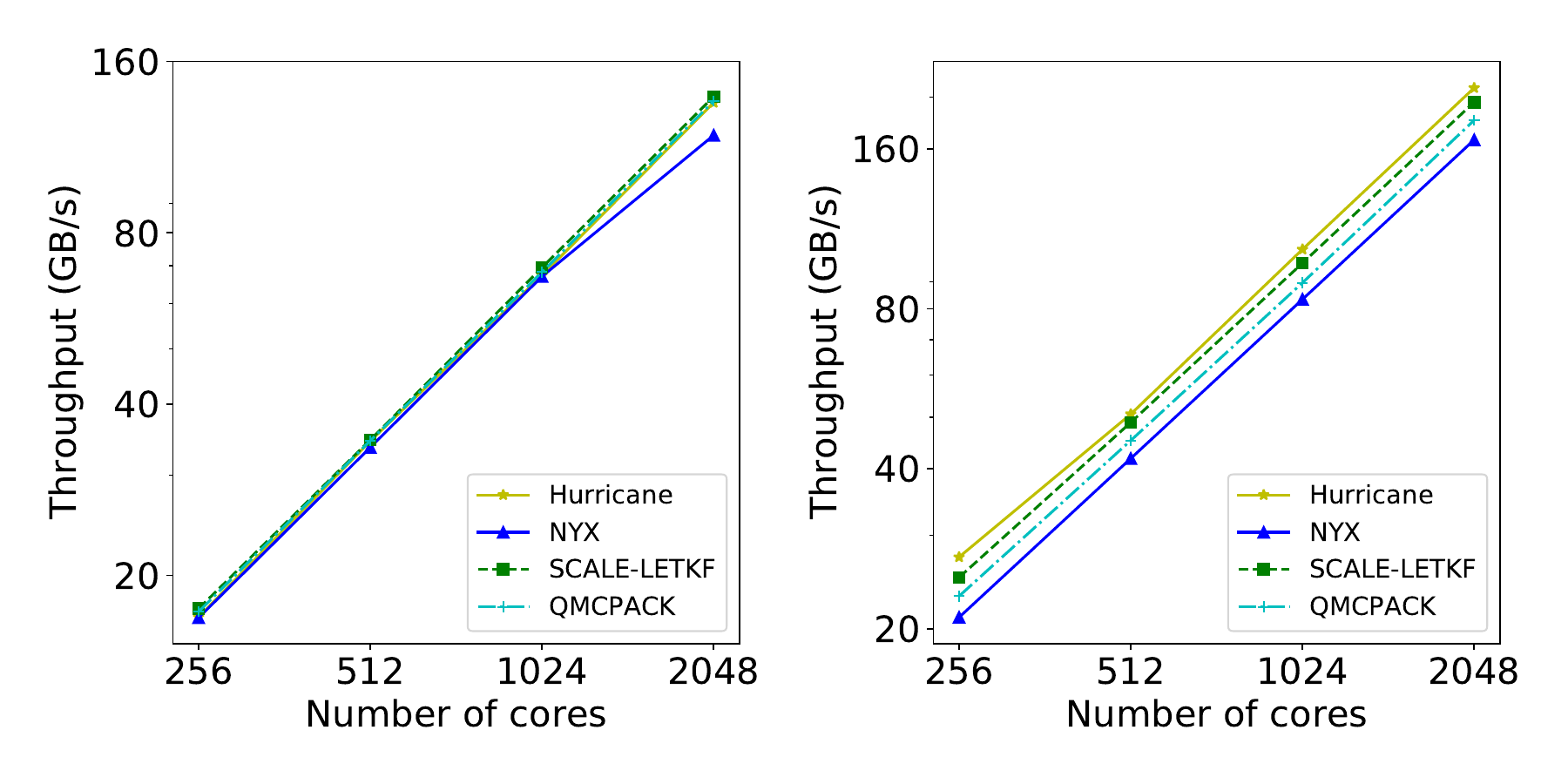}
\vspace{-6mm}
\caption{Scalability of the proposed compression method.}
\label{fig:scalability}
\end{figure}
\subsubsection{Scalability}
Because most data compression methods are designed in an embarrassingly parallel fashion, they are expected to have linear speedup when executed in parallel. 
We validate this by evaluating our method on 256, 512, 1024, and 2048 cores, respectively, with error bound $0.001$ for purposes of demonstration. 
This corresponds to 2.4~TB, 6~TB, 12.6~TB, and 1.2~TB of data with respect to the four datasets when 2k cores are used. 
As shown in Fig.~\ref{fig:scalability}, we observe almost linear speedup for both compression and decompression, which demonstrates the scalability of embarrassingly parallel data compression methods. 
Such characteristics promise high throughput when scale increases, which is very important for exascale data management.

\subsection{Compression quality}
We evaluate compression quality in terms of rate--distortion as introduced in Section~\ref{sec:formulation}. 
In what follows, we first evaluate the effect of the proposed techniques, namely level-wise quantization and adaptive decomposition, and compare our method with existing error-bounded lossy compressors.

\subsubsection{Impact of level-wise quantization and adaptive decomposition}
\begin{figure}[t] \centering
\vspace{-4mm}
\hspace{-4mm}
\includegraphics[width=0.5\textwidth]{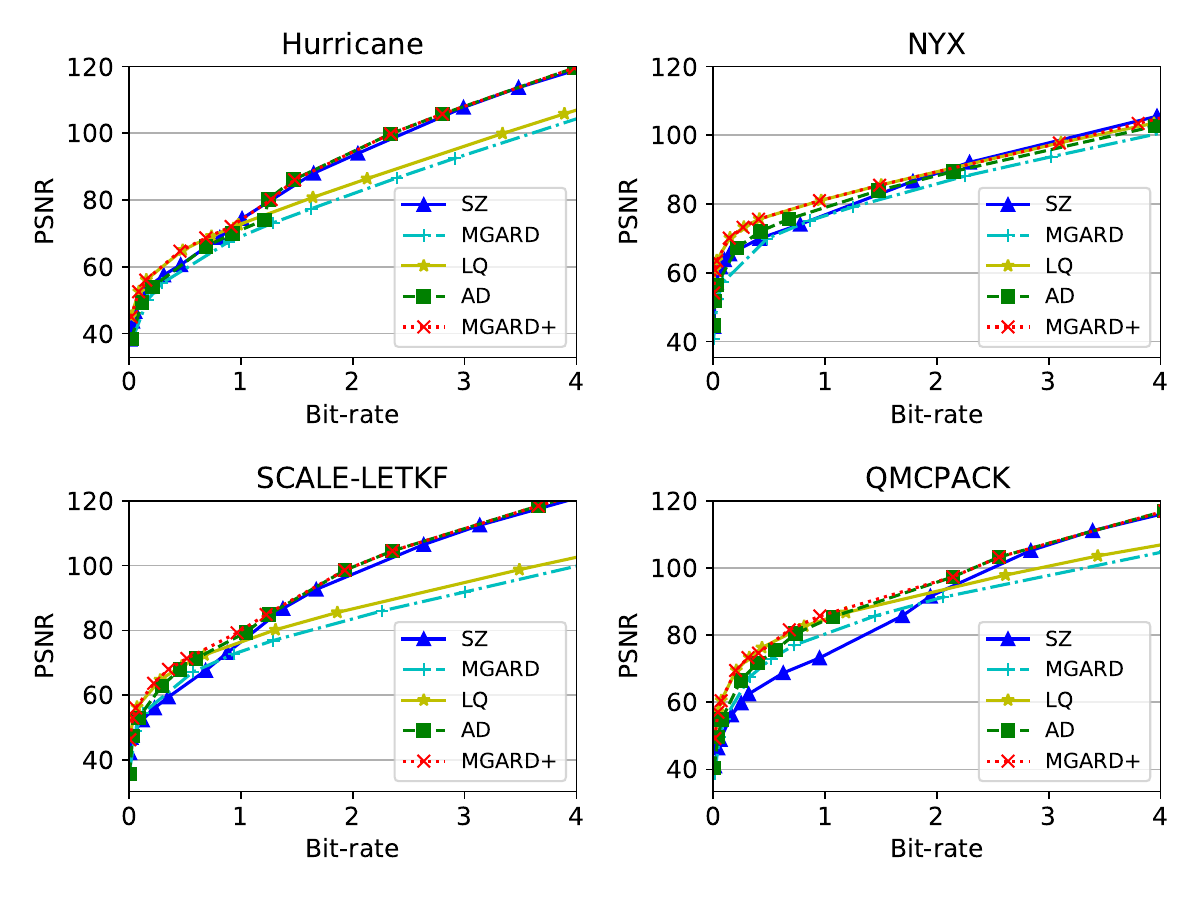}
\vspace{-6mm}
\caption{Impact of level-wise quantization and adaptive decomposition on rate--distortion.}
\label{fig:rate_distortion_opt}
\vspace{-4mm}
\end{figure}

We present the impact of the proposed techniques in Fig.~\ref{fig:rate_distortion_opt}.
The cyan line in this figure shows the rate--distortion curve of MGARD with uniform quantization across levels and extensive decomposition, which is the baseline. 
The yellow line (LQ) and green line (AD) display the rate--distortion curves after independently applying the level-wise quantization method proposed in Section~\ref{subsec:quantization} and the adaptive decomposition method introduced in Section~\ref{subsec:decomposition}, respectively. 
We also include the rate--distortion curve of SZ (the blue line) as another baseline for the adaptive decomposition method.
The red line illustrates the result of our method, which incorporates both level-wise quantization and adaptive decomposition.

From this figure, we can observe that both level-wise quantization (LQ) and adaptive decomposition (AD) have stable improvements over MGARD on all bit-rates with different emphasises. 
Specifically, level-wise quantization provides more advantages in small bit-rates (e.g., [0, 1]) while adaptive decomposition offers more improvements in large bit-rates (e.g., [1, 4]). 
This is consistent with our analysis in Section~\ref{subsec:decomposition} that the multilinear interpolation leveraged by the multilevel decomposition is more efficient for high error tolerance (i.e., small bit-rates) but less efficient for low error tolerance (i.e., large bit-rates) compared with the Lorenzo predictor used in SZ. 
With large bit-rates, when the Lorenzo predictor is always better, the approaches with adaptive decomposition degrade to SZ because they switch the prediction method at the first level. 
Our final solution, which integrates and takes advantages of both strategies, yields the best rate--distortion curve on all the four datasets.

\subsubsection{Comparison with state-of-the-art compressors}
\begin{figure}[t] \centering
\hspace{-4mm}
\includegraphics[width=0.5\textwidth]{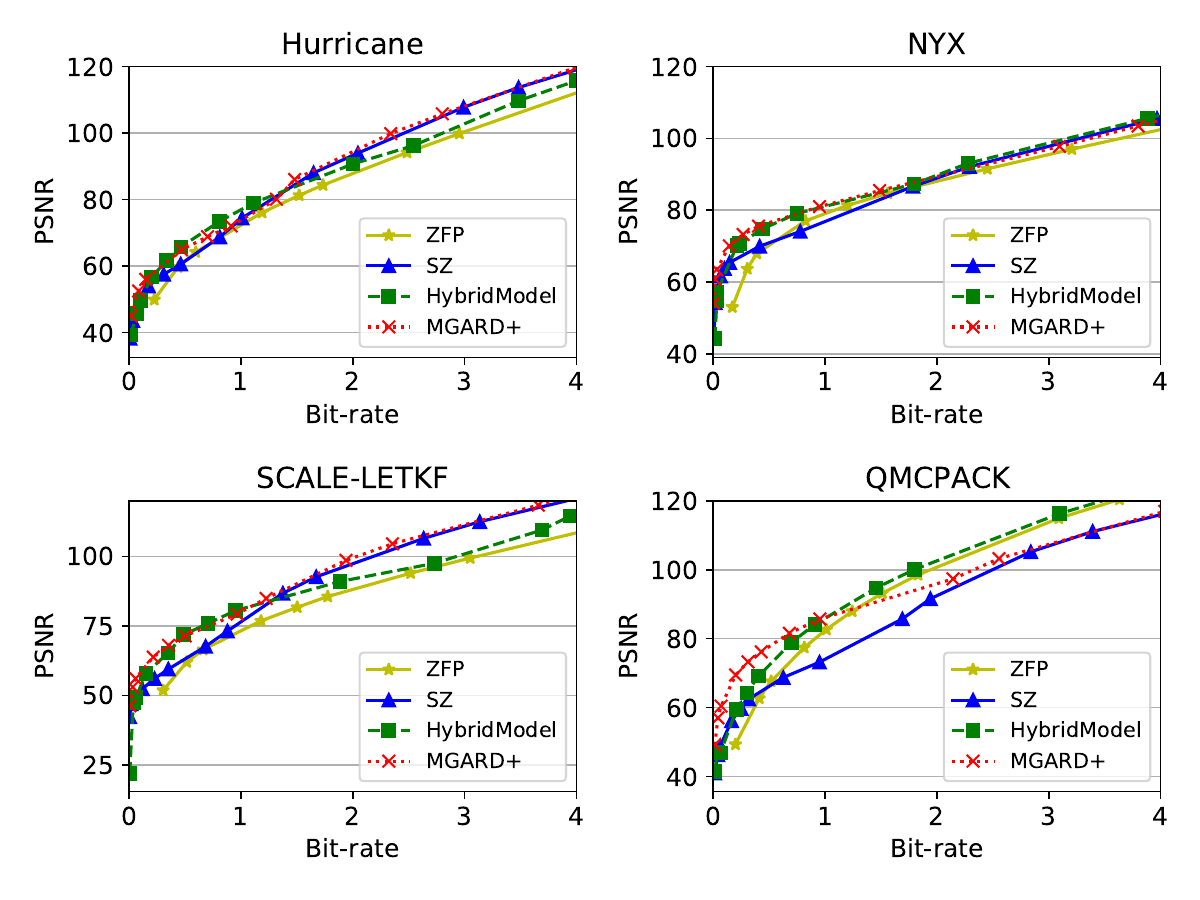}
\vspace{-6mm}
\caption{Rate--distortion curves of error-bounded lossy compressors on the four datasets (bit-rate $\in$ [0, 4]).}
\label{fig:rate_distortion_wide}
\vspace{-2mm}
\end{figure}

\begin{figure}[t] \centering
\vspace{-2mm}
\hspace{-4mm}
    \includegraphics[width=0.5\textwidth]{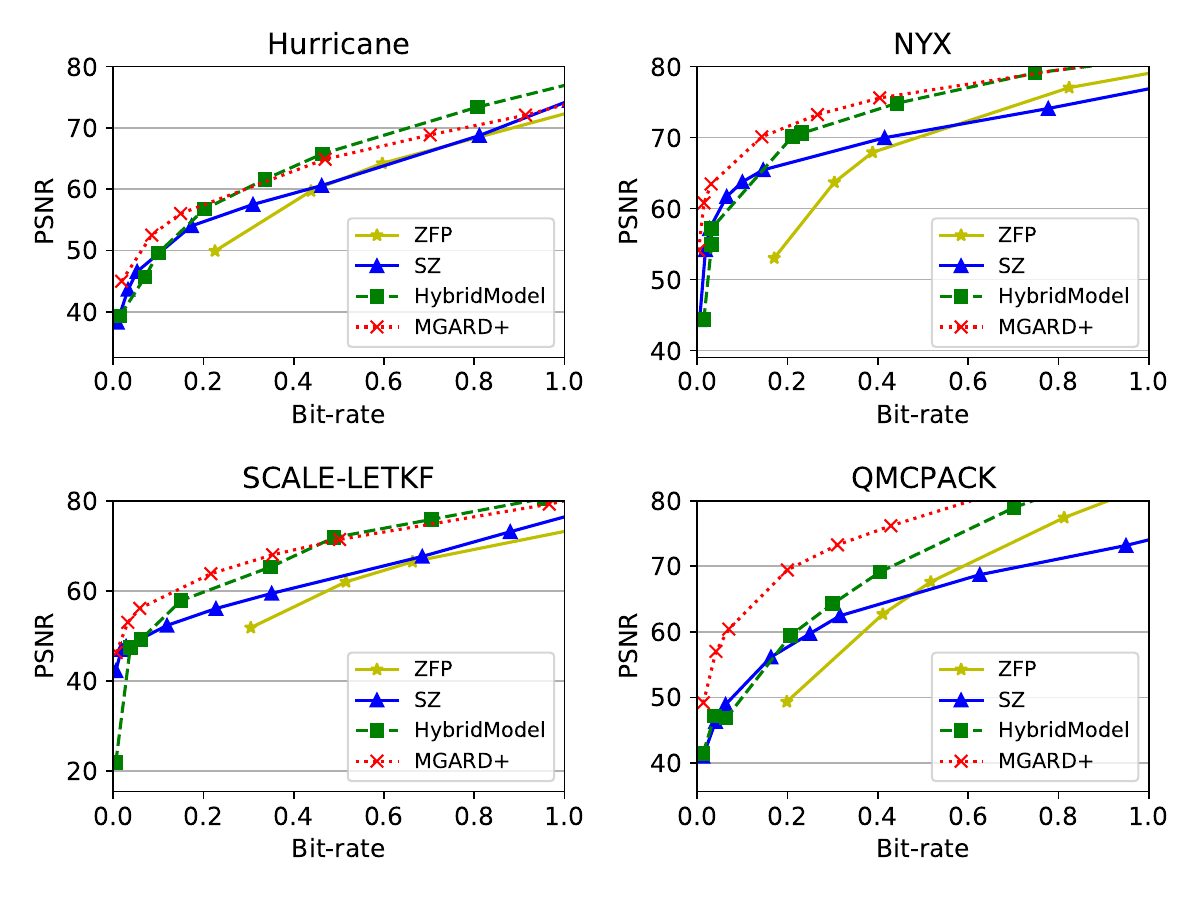}
    \vspace{-6mm}
\vspace{-1mm}
\caption{Rate--distortion curves of error-bounded lossy compressors on the four datasets (bit-rate $\in$ [0, 1]).}
\label{fig:rate_distortion}
\vspace{-3mm}
\end{figure}

We then compare the rate--distortion curves of our method with the other three state-of-the-art error-bounded lossy compressors in Fig.~\ref{fig:rate_distortion_wide} and Fig.~\ref{fig:rate_distortion}. 
We present the rate--distortion curve with bit-rate in $[0, 4]$, which corresponds to compression ratios $\geq 8$, and an enlarged view with bit-rate in $[0, 1]$, or equivalently compression ratios $\geq 32$. 
According to Fig.~\ref{fig:rate_distortion_wide}, our method leads to the least distortion at most bit-rates in most of the datasets.
One exception is QMCPACK with large bit-rates, where both the piecewise multilinear interpolation in the multilevel decomposition and the Lorenzo predictor in SZ are not as good as the transform-based de-correlation method used in ZFP and the hybrid model.
Nevertheless, our method can adapt to such scenarios as well by using either ZFP or the hybrid model as our external compressor in adaptive decomposition, which is our future work.
Compared with the other methods, our method improves the distortion in most cases for bit-rate range [0, 1] as shown in Fig.~\ref{fig:rate_distortion}, thanks to the robustness (against high error tolerance) of the piecewise multilinear interpolation used in the multilevel decomposition and the level-wise quantization method we proposed. 

\begin{table}[ht]
\vspace{-2mm}
\centering
\caption{Compression ratios (CR) and performance (Perf.) when PSNR $\approx$ 60.}
\label{tab:fixed_psnr_stat}
\begin{tabular}{|l|l|c|c|c|c|c|}
\hline
Datasets    & Compressors & PSNR & CR & Perf.\ (MB/s)\\
\hline
\multirow{4}{*}{Hurricane} & SZ  & 59.97 & 74.08 & 105.02 \\
\cline{2-5}
& ZFP & 59.68 & 73.01 & 335.63 \\ 
\cline{2-5}
& HybridModel & 59.99 & 110.66 & 43.73\\ 
\cline{2-5}
& MGARD+ & 60.21 & \textbf{120.08} & 96.03\\
\hline
\multirow{4}{*}{NYX} & SZ  & 59.84 & 722.72 & 115.84\\
\cline{2-5}
& ZFP & 59.44 & 130.44 & 346.47 \\ 
\cline{2-5}
& HybridModel & 59.20 & 834.49 & 87.95\\ 
\cline{2-5}
& MGARD+ & 60.12 & \textbf{2525.93} & 94.23\\
\hline
\multirow{4}{*}{SCALE-LETKF} & SZ  & 59.42 & 91.1 & 108.22\\
\cline{2-5}
& ZFP & 57.49 & 79.14 & 287.78\\ 
\cline{2-5}
& HybridModel & 59.38 & 176.38 & 57.78\\ 
\cline{2-5}
& MGARD+ & 59.82 & \textbf{252.53} & 104.82\\
\hline
\multirow{4}{*}{QMCPACK} & SZ  & 59.73 & 128.26 & 106.11\\
\cline{2-5}
& ZFP & 57.76 & 99.08 & 291.72\\ 
\cline{2-5}
& HybridModel & 59.53 & 148.09 & 39.70\\ 
\cline{2-5}
& MGARD+ & 60.42 & \textbf{467.85} & 100.30\\
\hline
\end{tabular}
\end{table}

\begin{figure}[ht] \centering
\hspace{-4mm}
\subfigure[{Original}]
{
\includegraphics[width=0.32\columnwidth]{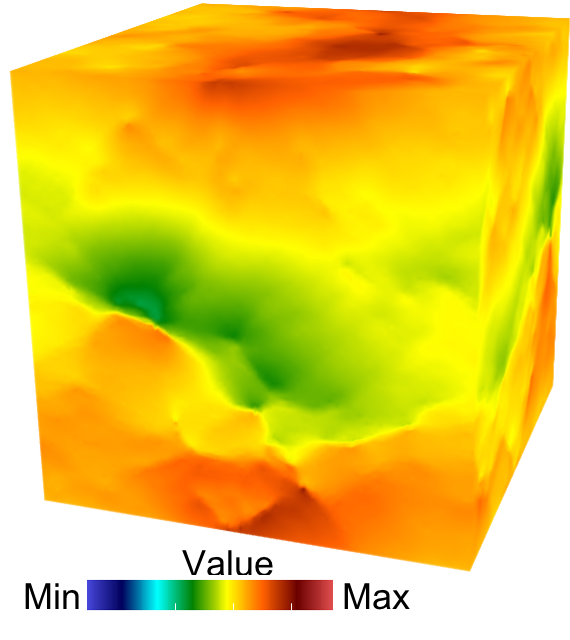}
}
\hspace{-2mm}
\subfigure[{Decompressed}]
{
\includegraphics[width=0.32\columnwidth]{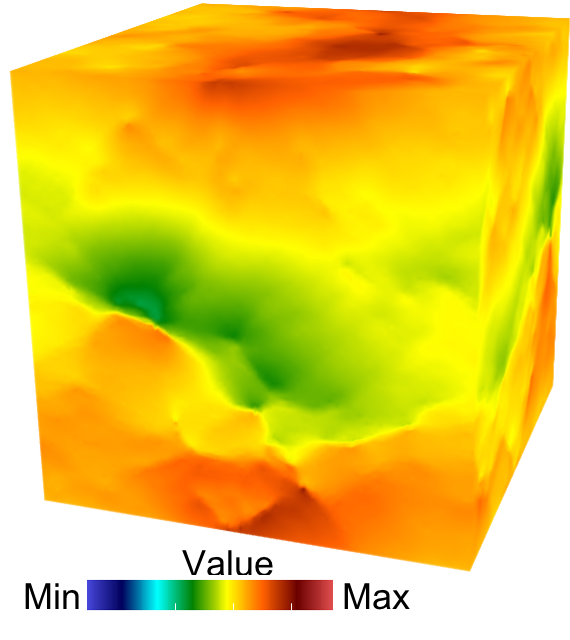}
}
\hspace{-2mm}
\subfigure[{Relative error}]
{
\includegraphics[width=0.32\columnwidth]{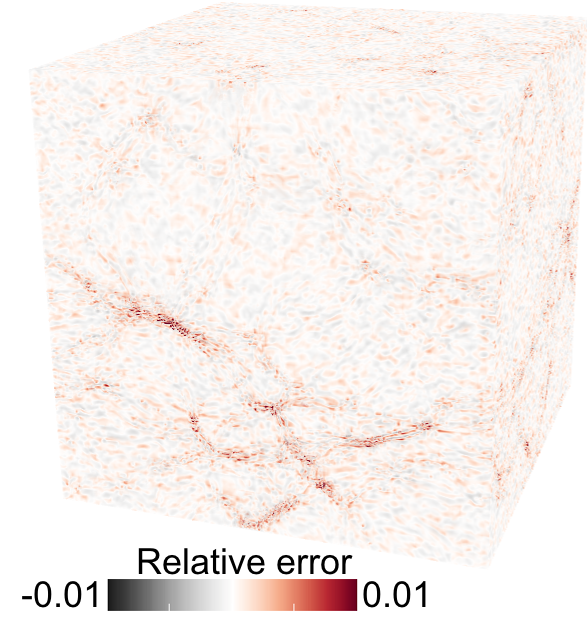}
}
\hspace{-4mm}
\vspace{-2mm}
\caption{Visualization of original data, decompressed data, and relative error of NYX velocity\_x field using our method (dataset PSNR = 60.12, compression ratio = 2525, field PSNR = 57.37, compression ratio = 1396).}
\label{fig:visualization}
\vspace{-4mm}
\end{figure}

We further show the compression ratios and performance of the evaluated error-bounded lossy compressors on the four datasets when tuning them to have almost the same distortion in Table~\ref{tab:fixed_psnr_stat}. 
We use a PSNR of around $60$ for demonstration purposes, because such PSNR is able to provide valid data for visualization purposes, as displayed in Fig.~\ref{fig:visualization}.
This figure visualizes the original data of NYX velocity\_x field, as well as the decompressed data using our compression method, which shows almost no visual difference even under such a high compression ratio.
Although the throughput of our method is slower than that of ZFP, it offers $2\times \sim 20\times$ improvements on the compression ratios under the same distortion. 
Also, our method has similar compression performance compared to that of SZ, which is almost $2\times$ as fast as the hybrid model in most cases. 
Generally speaking, our method has up to $2\times$ compression improvement over that of the best existing methods, which could be a very good option to reduce the storage requirement and I/O intensity for exascale systems.


\section{Related Works} \label{sec:related}
Storage limitations and I/O bottlenecks have become serious problems for large-scale scientific applications. Data compression is a direct way to address such problems, and many approaches have been proposed in literature.

Lossless compressors~\cite{alted2017blosc, fpc, gzip, zstd} are developed to recover exact data while reducing the size, but they only achieve limited compression ratios for floating-point scientific data. 
According to recent studies~\cite{lindstromerror}, their compression ratios are usually less than $3$, which is insufficient for today's large-scale scientific simulations and experimental devices. 
General lossy compressors~\cite{jpeg, jpeg2000, li2019vapor} are able to provide decent compression ratios.
But they are not preferred for scientific data analysis because they do not respect specific error requirements from application users.

Error-bounded lossy compression~\cite{sz16, zfp, fpzip, isabela, numarck, sz-reg, sz-pwr, sz-hybrid} is proposed to trade off data accuracy for compression ratio, which features in offering high compression ratios while controlling data distortion.
General error-bounded lossy compressors can be divided into two categories, namely prediction-based ones and transform-based ones, based on how they decorrelate original data. 
Prediction-based lossy compressors such as~\cite{sz16, sz-reg, sz-hybrid} leverage prediction models to decorrelate original data, while transform-based ones such as~\cite{zfp} rely on invertible transforms to do so.
According to previous work~\cite{understand-compression-ipdps18}, SZ~\cite{sz-reg} and ZFP~\cite{zfp} are the two best error-bounded lossy compressors of their kinds. 
They usually lead to higher compression quality under the same distortion when compared with other approaches. 
As a multi-algorithm prediction-based compressor, SZ decomposes data into small blocks (e.g., $6\times6\times6$ blocks for 3D datasets), and adaptively selects the best-fit prediction method between the Lorenzo predictor and a linear-regression based predictor for data in each block. 
The prediction differences are then fed to a pipeline of linear-scaling quantization, customized variable-length encoding, and lossless compression to generate the compressed byte streams. 
ZFP, a transform-based lossy compressor, processes data in $4^d$ blocks following the order of exponent alignment, fixed-point alignment, nonorthogonal transform, and embedded encoding. 
Generally speaking, none of these compressors can be always better than the others according to the literature~\cite{sz-hybrid}.

Attempts have been made to combine SZ and ZFP for better compression ratios under the same distortion in terms of PSNR.
A previous approach~\cite{sz-hybrid} proposed to use the nonorthogonal transform in ZFP as a predictor in SZ, but it suffered from high overhead in terms of performance because of a costly iterative sampling strategy for best-fit predictor selection. 
Another approach~\cite{sz-aos} tried to select the better one between the best of SZ and ZFP, but it at most provides the same compression quality as either SZ or ZFP.

Recently, multilevel data reduction~\cite{ainsworth2018multilevel, ainsworth2019multilevel, ainsworth2019multilevelerror} has been proposed by the applied math community for error-bounded lossy compression. 
However, such algorithm is not tailored for both reduction performance (throughput) and quality (compression ratios under a given distortion). 
In this work, we propose two novel techniques to improve the compression ratios of the multilevel data reduction algorithms under the same PSNR, along with a series of optimizations to achieve high compression/decompression throughput.
Our level-wise quantization strategy accounts for the different impacts of errors in each level, and the adaptive decomposition strategy is, to the best of our knowledge, the first to combine multilevel method with other compressors to form a new error-bounded lossy compressor.
\section{Conclusion}\label{sec:conclusion}
In this paper, we present two novel techniques to enhance the compression quality of multilevel data reduction, as well as a series of optimizations to improve its performance. 
The proposed approach leads to up to $2\times$ compression ratio gain compared to state-of-the-art error-controlled lossy compressors under the same distortion and tens of performance improvement over the existing multilevel method. 
In future work, we plan to further improve the quality of multilevel data reduction by exploring higher-order basis functions and adapting them in different regions of data.


%



\ifCLASSOPTIONcompsoc
  \section*{Acknowledgments}
\else
  \section*{Acknowledgment}
\fi

This research was supported by the Exascale Computing Project (17-SC-20-SC), a collaborative effort of U.S. Department of Energy Office of Science and the National Nuclear Security Administration. This material is also based upon work supported by the U.S. Department of Energy, Office of Science, Office of Advanced Scientific Computing Research (ASCR), Scientific Discovery through Advanced Computing (SciDAC) program. This research used resources of the Oak Ridge Leadership Computing Facility, which is a DOE Office of Science User Facility.

\ifCLASSOPTIONcaptionsoff
  \newpage
\fi



%

\bibliographystyle{IEEEtran}
\bibliography{bib/references}



%


\cleardoublepage
\begin{IEEEbiography}[{\includegraphics[width=1in,height=1.25in,clip,keepaspectratio]{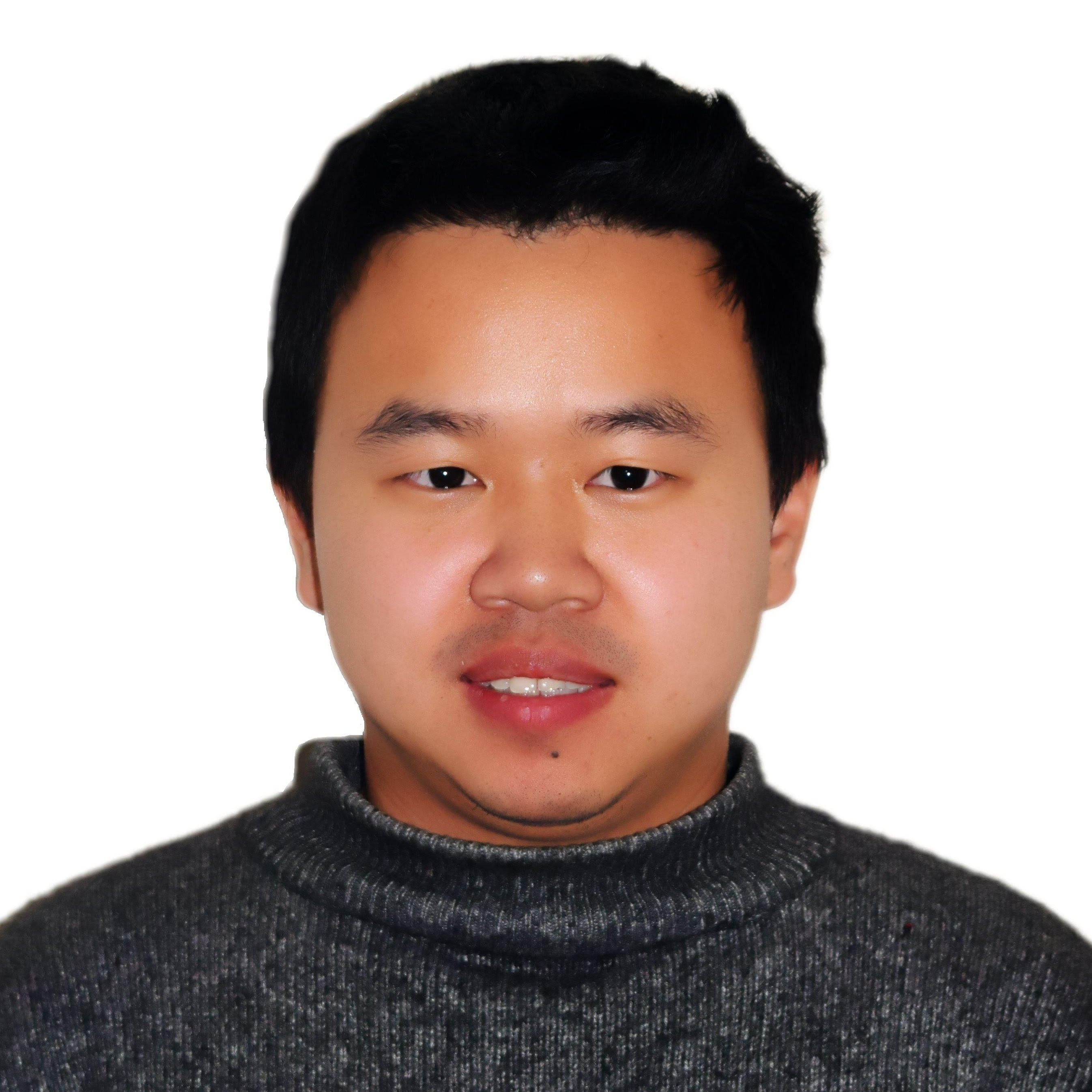}}]{Xin Liang} 
is a Computer/Data Scientist in the Workflow Systems Group at Oak Ridge National Laboratory. He received his Ph.D. degree from University of California, Riverside in 2019 and his bachelor's degree from Peking University in 2014. His research interests include high-performance computing, parallel and distributed systems, data management and reduction, big data analytic, scientific visualization, and cloud computing. He is a member of the IEEE. 
\end{IEEEbiography}
\vspace{-6.5mm}
\begin{IEEEbiography}[{\includegraphics[width=1in,height=1.25in,clip,keepaspectratio]{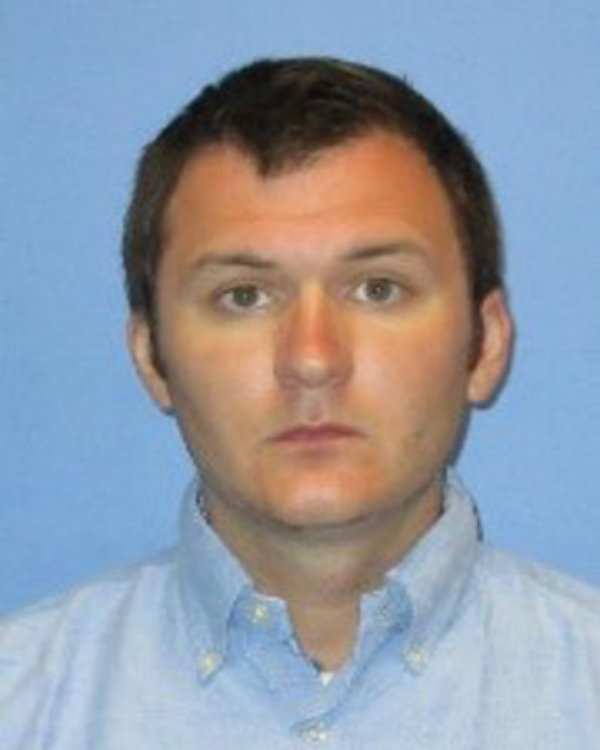}}]{Ben Whitney}
is a postdoctoral research associate in the Computer Science and Mathematics Division at Oak Ridge National Laboratory (ORNL).
He received his Ph.D. degree in applied mathematics from Brown University in 2018.
His research interests include data compression, numerical methods, and scientific software development.
\end{IEEEbiography}
\vspace{-6.5mm}
\begin{IEEEbiography}[{\includegraphics[width=1in,height=1.25in,clip,keepaspectratio]{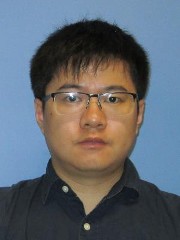}}]{Jieyang Chen} 
is a Computer Scientist in the Computer Science and Mathematics Division at Oak Ridge National Laboratory. He received his master and Ph.D. degrees in Computer Science from University of California, Riverside in 2014 and 2019. He received a bachelor's degree in Computer Science and Engineering from Beijing University of Technology in 2012. 
His research interests include high-performance computing, parallel and distributed systems, and big data analytics. 
He is a member of the IEEE.
\end{IEEEbiography}
\vspace{-6.5mm}
\begin{IEEEbiography}[{\includegraphics[trim={5mm 0 5mm 0}, width=1in,height=1.25in,clip,keepaspectratio]{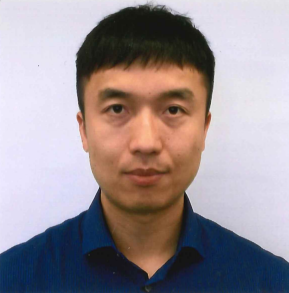}}]{Lipeng Wan}
is a Computer Scientist in the Computer Science and Mathematics Division at Oak Ridge National Laboratory (ORNL). He received his Ph.D. degree in computer science from the University of Tennessee, Knoxville in 2016. Prior to that, he earned his master degree from Southeast University, China, in March 2011, and his bachelor degree from Nanjing University of Science and Technology, China, in June 2008. His research mainly focuses on scientific data management and high-performance computing.
\end{IEEEbiography}
\vspace{-6.5mm}
\begin{IEEEbiography}[{\includegraphics[width=1in,height=1.25in,clip,keepaspectratio]{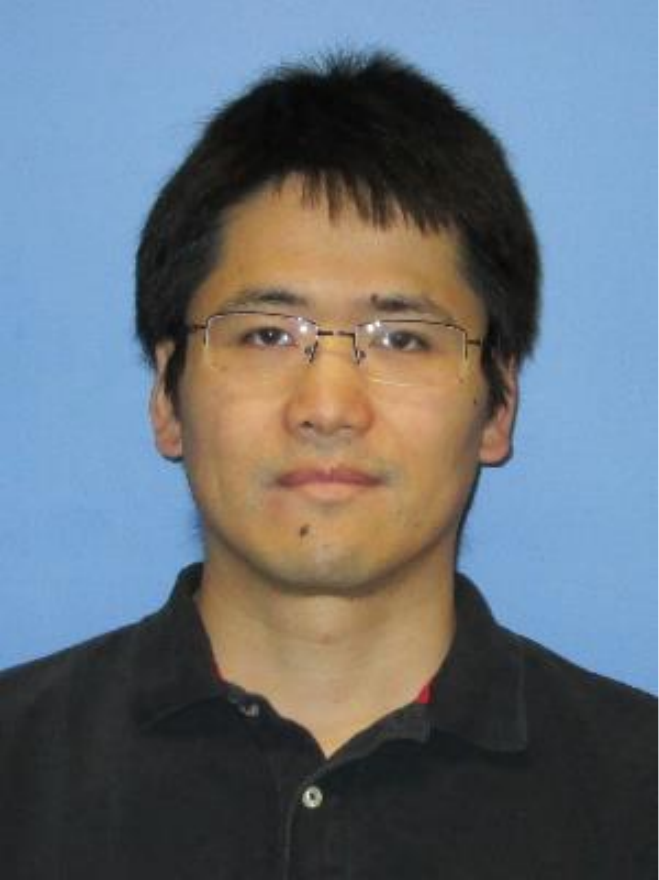}}]{Qing Liu}
is an Assistant Professor in the Department of Electrical and Computer Engineering at NJIT. He has joint faculty appointment with Oak Ridge National Laboratory. Prior to that, he was a staff scientist at Science Data Group, Oak Ridge National Laboratory. He received his Ph.D. in Computer Engineering from the University of New Mexico in 2008. His areas of interest include high-performance computing, data science, and networking. In 2013 He won an R\&D 100 award for the development of the Adaptable I/O System for Big Data.
\end{IEEEbiography}
\vspace{-4mm}
\begin{IEEEbiography}[{\includegraphics[width=1in,height=1.25in,clip,keepaspectratio]{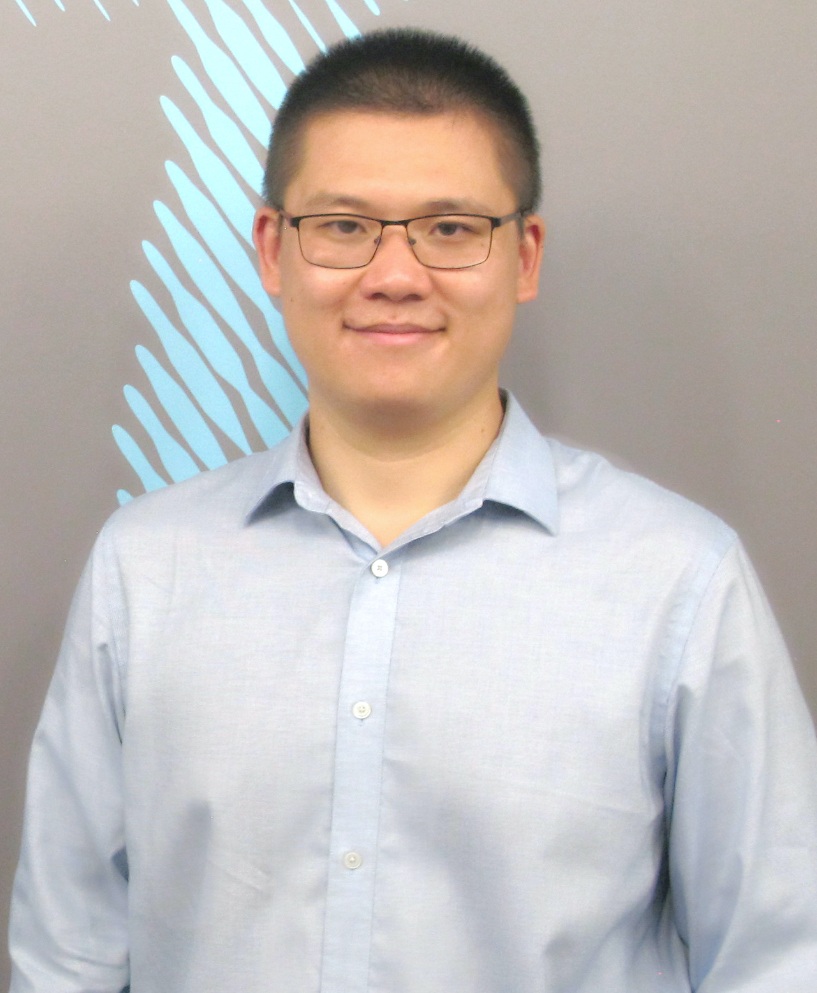}}]{Dingwen Tao}
is an assistant professor in of computer science at Washington State University. He received his Ph.D. in Computer Science from University of California, Riverside in 2018 and B.S. in Mathematics from University of Science and Technology of China in 2013. He works at the intersection of HPC and big data analytics, focusing on scientific data management, HPC storage and I/O, fault tolerance at extreme scale, and distributed machine learning. 
He was the receipt of the 2020 IEEE Computer Society TCHPC Early Career Researchers Award for Excellence in HPC.
\end{IEEEbiography}
\begin{IEEEbiography}[{\includegraphics[width=1in,height=1.25in,clip,keepaspectratio]{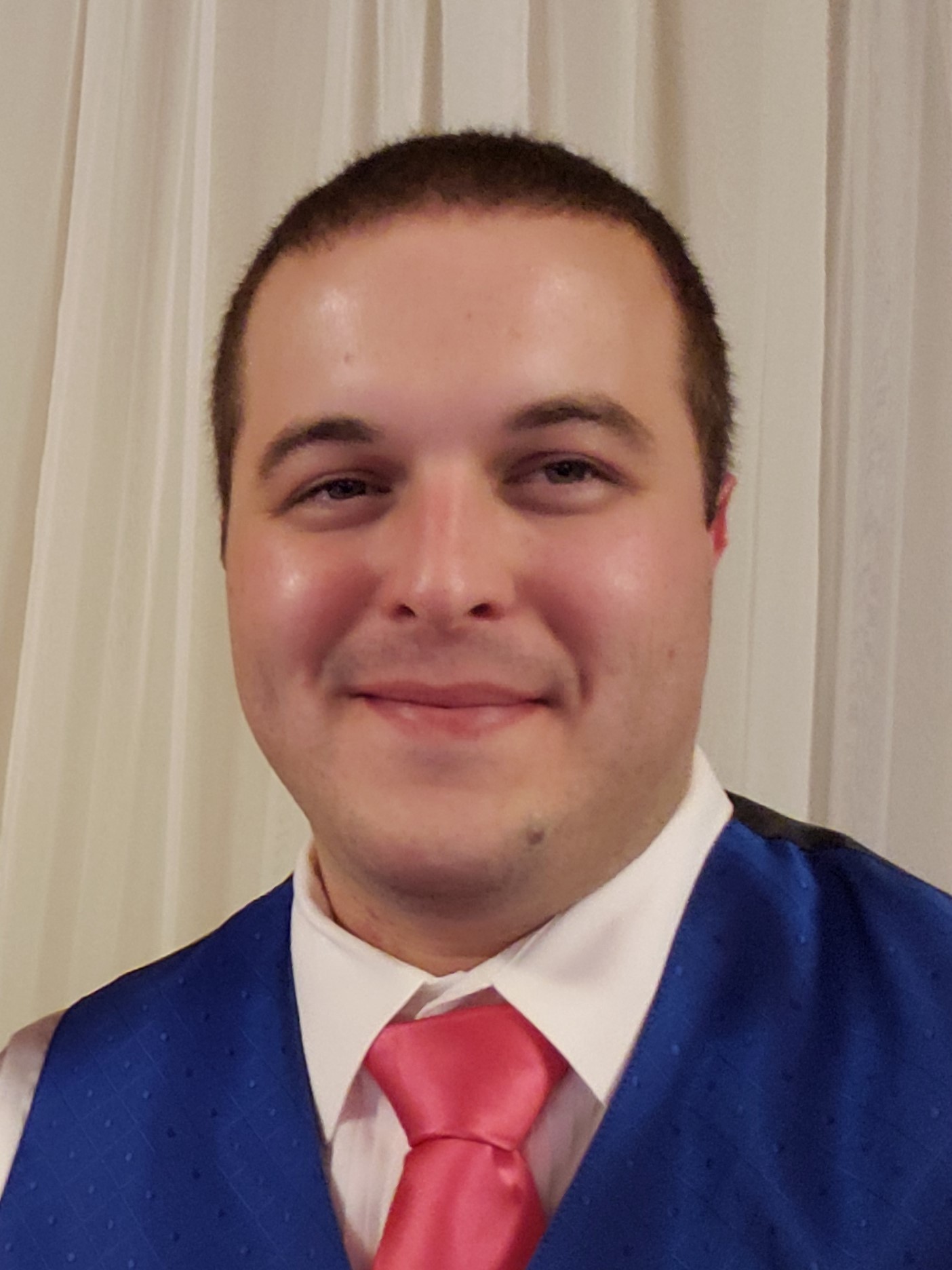}}]{James Kress}
is a Research Scientist in scientific visualization in the Data and AI Section at Oak Ridge National Laboratory. He received his PhD in 2020 and his Master's in 2017 from the University of Oregon in Computer Science. He received his BS in 2013 from Boise State University in Computer Science, with a minor in Political Science. His research is focused on in situ scientific visualization, HPC, and the intersection of the two. He has published in high-quality conferences and journals, including ISC, TVCG, ICDCS, and IEEE CGA. 
\end{IEEEbiography}
\vspace{-6.5mm}
\begin{IEEEbiography}[{\includegraphics[width=1in,height=1.25in,clip,keepaspectratio]{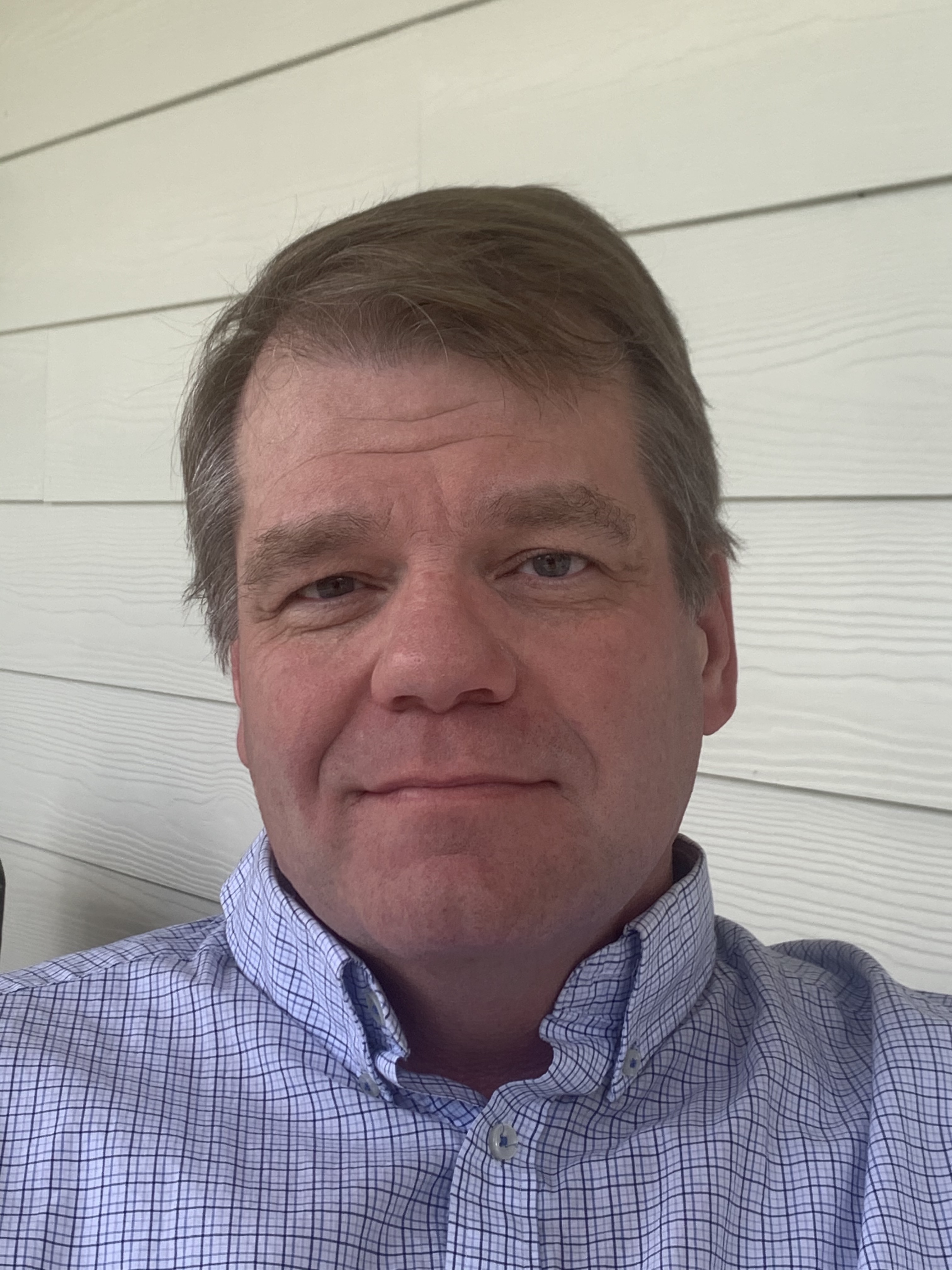}}]{David Pugmire}
is a Senior Research Scientist and Visualization Group Leader in the Data and AI Section at Oak Ridge National Laboratory (ORNL) and Joint Faculty Professor at the University of Tennessee. He received his Ph.D. from the University of Utah in 2000. Before joining ORNL, he was a Research Scientist at Los Alamos National Laboratory. His research interests include scalable visualization on high performance computing systems. In 2006 he won an R\&D 100 award for an NPU-based image compositor. He has published numerous papers in high-quality conferences and journals, including SC, IEEE Visualization, EuroVis, IPDPS, BigData and TVCG. He is a member of the ACM.
\end{IEEEbiography}
\vspace{-6.5mm}
\begin{IEEEbiography}[{\includegraphics[width=1in,height=1.25in,clip,keepaspectratio]{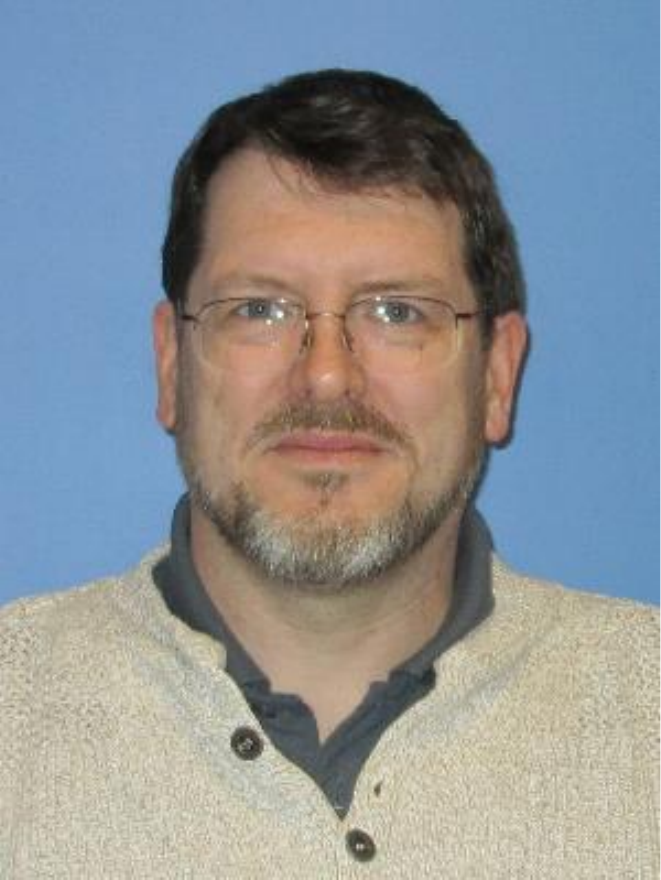}}]{Matthew Wolf}
is a Senior Computer Scientist in the Computer Science and Mathematics Division at Oak Ridge National Laboratory (ORNL).  Before joining ORNL full-time, he was a senior research scientist at Georgia Institute of Technology with a joint appointment to ORNL.  His research interests include high performance computing, adaptive I/O and messaging middleware, and in situ analysis and visualization systems for science.  He has advised and co-advised numerous students who have gone on to careers in industry, academia, and the national laboratories, as well as publishing papers in prominent conferences and journals. 
\end{IEEEbiography}
\vspace{-6.5mm}
\begin{IEEEbiography}[{\includegraphics[width=1in,height=1.25in,clip,keepaspectratio]{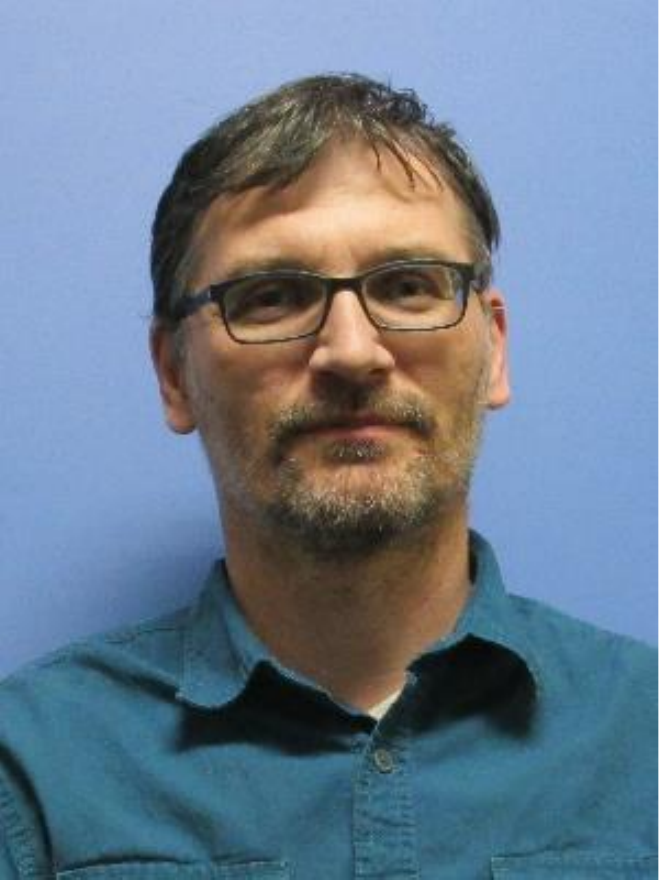}}]{Norbert Podhorszki}
Norbert Podhorszki is a Senior Research Scientist in the Workflow Systems Group at the Oak Ridge National Laboratory. He is one of the key developers of ADIOS that won an R\&D100 award in 2013. His main research interest is in creating I/O and staging solutions for in-situ processing of data on leadership class computing systems. He received his Ph.D. in Information Technology from the Eötvös Loránd University of Budapest. He has worked in the field of logic programming, performance monitoring and analysis of message-passing programs, application monitoring in Grid environments and application development in
Desktop Grids, scientific workflow technologies in supercomputing projects, and high performance I/O.
\end{IEEEbiography}
\vspace{-6.5mm}
\begin{IEEEbiography}[{\includegraphics[width=1in,height=1.25in,clip,keepaspectratio]{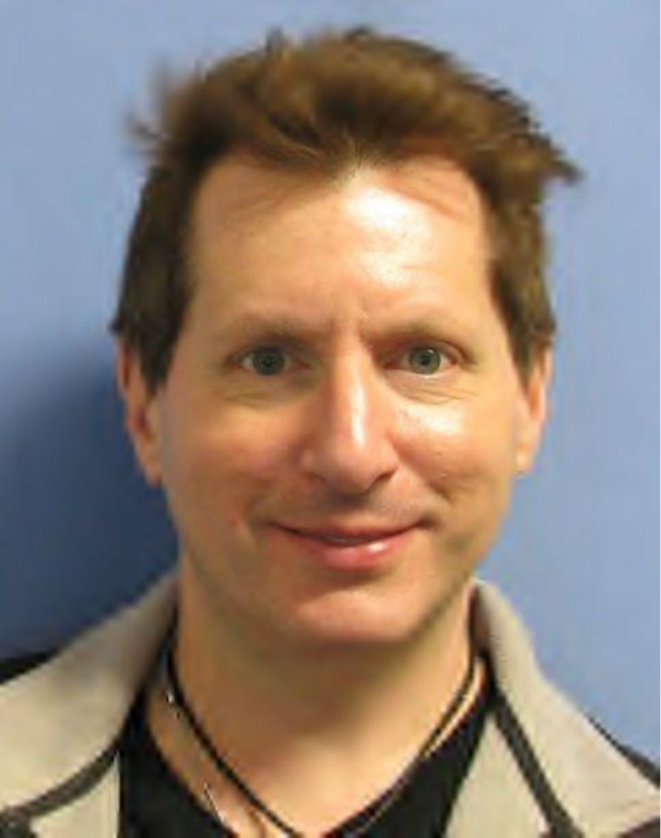}}]{Scott Klasky}
is a distinguished scientist and Group Leader in the Computer Science and Mathematics Division at the Oak Ridge National Laboratory (ORNL). He also
has a joint faculty appointment at the University of Tennessee, Knoxville, and an adjunct position at Georgia Tech.  Prior to that, he was a 
staff scientist at the Princeton Plasma Physics Laboratory, a Senior
Scientist at Syracuse University and a Post Doc at the University of
Texas at Austin. He received his Ph.D. in Physics from the University
of Texas at Austin. He is a senior Member of IEEE, and won an R\& D 100
award for being the leader for the Adaptable I/O System (ADIOS). 
He has almost 300 publications in the fields of computer science,
data management, storage and I/O, workflow automation, data reduction,
data visualization and physics.
\end{IEEEbiography}





\end{document}